\documentclass[aps,pre,reprint,showpacs,eqsecnum]{revtex4-1}
\usepackage{amssymb,amsmath}
\usepackage{graphicx}
\usepackage{color}

\newcommand{\mn}[1]{{\mathbf #1}}
\newcommand{\thr}{\text{th}}
\newcommand{\hnn}[1]{\widehat{\boldsymbol #1}}
\newcommand{\medio}[1]{\langle #1\rangle}

\newcommand\beq{\begin{equation}}
\newcommand\eeq{\end{equation}}
\newcommand\beqa{\begin{eqnarray}}
\newcommand\eeqa{\end{eqnarray}}
\def\bal#1\eal{\begin{align}#1\end{align}}
\newcommand{\nn}{\nonumber\\}

\newcommand\qq{\gamma}
\newcommand\cum{c}

\begin{document}
\title{Hydrodynamic Burnett equations for inelastic Maxwell models {of granular gases}}
\author{Nagi Khalil}
\author{Vicente Garz\'{o}}
\author{Andr\'{e}s Santos}
\affiliation{Departamento de F\'{\i}sica, Universidad de Extremadura, E-06071 Badajoz, Spain}


\begin{abstract}
The hydrodynamic Burnett equations and the associated transport coefficients are exactly evaluated for  generalized inelastic Maxwell models. In those models, the one-particle distribution function obeys the inelastic Boltzmann equation, with a velocity-independent  collision rate  proportional to  the $\qq$ power of the temperature.  The pressure tensor and the heat flux are obtained to second order in the spatial gradients of the hydrodynamic fields with explicit expressions for all the Burnett transport coefficients  as  functions of $\qq$, the coefficient of normal restitution, and the dimensionality of the system. Some transport coefficients that are  related in a simple way in the elastic limit become decoupled in the inelastic case. As a byproduct, existing results in the literature for three-dimensional elastic systems are recovered, and a generalization to any dimension of the system is given.
The structure of the present results is used to estimate the Burnett coefficients for inelastic hard spheres.
\end{abstract}

\date{\today}
\maketitle

\section{Introduction}
\label{sec1}

Kinetic theory provides a fundamental and systematic way of deriving closed hydrodynamic equations for dilute molecular gases by means of the Chapman--Enskog (CE) method \cite{CC70}. The essential ingredients of the method are a Boltzmann-like kinetic equation for the distribution function, an identification of the hydrodynamics fields, and an expansion in powers of gradients of those hydrodynamic fields \cite{CC70,RL77}. The generality of the above scheme allows for the use of the CE method in the study of different systems, the reliability of the resulting description being dependent, on the one hand, on the validity of the kinetic equation used and the choice of the hydrodynamic variables and, on the other hand, on the fulfillment of the hypothesis of weak spatial gradients.

For granular fluids, which can be briefly defined as systems composed by macroscopic particles with  short-ranged inelastic interactions (collisions), a closed hydrodynamic description based on the CE method has been derived for different models. Two of them are relevant here, the inelastic hard-sphere model (IHSM) and the inelastic Maxwell model (IMM) \cite{BCG00,BK00,EB02b,BK03,GS11}. The minimal version of the IHSM corresponds to a collection of smooth hard spheres or disks that undergo inelastic collisions, with a velocity-independent coefficient of normal restitution $\alpha$ \cite{G03,BP04}. More sophisticated models, close to the IHSM, consider particle rotations with coefficients of normal and tangential restitution \cite{JR85,ZTPSH98,GNB05,Z06,BPKZ07,SKG10,SKS11}, velocity-dependent coefficients of restitution \cite{BP01,BP03,BP04}, polydispersity \cite{G08b}, presence of an interstitial fluid \cite{KH01,GTSH12,GCV13,KG13}, etc. Some conclusions of the research carried out along the last few years in the minimal version of the IHSM, and also in some others, are that the inelastic Boltzmann  equation is able to describe dilute (and moderately dense) systems (the fundamental hydrodynamic variables being  the same as that of the ordinary elastic case, i.e., density, velocity, and  temperature) and the Navier--Stokes (NS) hydrodynamic equations provided by the CE method are applicable for a generality of accessible situations with small spatial gradients. Therefore, the current attempts to extend the NS hydrodynamic description for dilute granular gases \cite{BDKS98,GD02} focus on several fronts:  denser regimes \cite{GD99,GDH07,GHD07}, even taking into account velocity correlations \cite{vNEB98}, inclusion of non-Newtonian states like the uniform shear flow \cite{BRM97,SGD04,L06,G06,VSG11}, the Fourier state \cite{BCMR01,BKR09}, and high gradients \cite{SG98}.

The latter limitation (i.e., the weakness of the spatial gradients) of the usual hydrodynamic description is addressed in this work.
More specifically, we apply the CE method to the next order to the NS one, namely the Burnett order, where the irreversible momentum and heat fluxes are obtained to second order in the hydrodynamic gradients.
The importance of going beyond the NS order in granular gases, due to the inherent coupling between inelasticity and spatial gradients, was pointed out by pioneering works a few years ago \cite{SG98,G08a}.
On the other hand, the derivation of the Burnett equations in the framework of the Boltzmann equation for the IHSM is an extremely difficult task that requires the use of approximations to get high-degree collisional moments. In fact, to the best of our knowledge, the existing Burnett hydrodynamics description of the IHSM \cite{SG98} makes use of the elastic forms of the Burnett transport coefficients \cite{CC70}.

A way of circumventing the above difficulty, while keeping the structure of the nonlinear Boltzmann equation, consists of using the IMM, where calculations can be made exactly  for any degree of dissipation.
In this model, the  collision rate of the inelastic Boltzmann equation is assumed to be  independent of the relative velocity of the colliding particles, just as in the case of elastic collisions  \cite{TM80,GS03}. Furthermore, in order to capture in an effective way the velocity dependence of the original IHSM collision rate, one usually assumes that the IMM collision rate is proportional to $T^\qq$ with $\qq=\frac{1}{2}$, where $T$ is the local granular temperature. In this paper, we take $\qq$ as a generalized exponent, so that different values of $\qq$ can be used to mimic different interactions. For instance, in the case of elastic collisions, a repulsive potential of the form $\phi(r)\sim r^{-s}$ corresponds to $\qq=1/2-(d-1)/s$, where $d$ is the dimensionality of the system \cite{E81}, so that $\qq=0$ defines the standard Maxwell model [$s=2(d-1)$], while $\qq=\frac{1}{2}$ mimics hard spheres ($s\to\infty$).

The derivation of the Burnett equations for the IMM  can be essentially done thanks to the exact knowledge of the collisional moments up to fourth degree for arbitrary values of the coefficient of restitution and the dimensionality of the system \cite{GS07}.
The price paid for obtaining exact results  is to have a less realistic description than with the IHSM. Nevertheless, it has been shown that the transport properties obtained from the IMM compare quite well with those of the IHSM \cite{S03,G03a,GA05,G07,SG07}. Moreover, experiments for magnetic grains can be well described by means of the IMM \cite{KSSAOB05}.  In addition,  the structure of the Burnett constitutive equations of the IMM are expected  to be the same as  that of IHSM.  Apart from that, the results of the present work have their own  interest since they constitute a natural extension of the Burnett  hydrodynamic description of Maxwell molecules \cite{CC70} to granular gases. As we will see, some Burnett transport coefficients having simple relationships in the elastic limit decouple in the inelastic case.

{While the knowledge of the Burnett constitutive equations can be useful for the description of non-Newtonian granular flows, some caution is required because, as reported for ordinary gases in Bobylev's pioneering work \cite{B82}, the Burnett hydrodynamic equations are \emph{artificially} unstable. On the other hand, several methods of regularization of the Burnett equations have been proposed to overcome the above difficulty \cite{UVG00,JS01,B04,B06,CKK07,G08a}. In principle, those methods could be applied to the inelastic case in order to disentangle Bobylev's instability from the   clustering instability that can be present in granular gases \cite{GZ93,M93b}.}

This work is organized as follows. In Sec.\ \ref{sec2}, the general CE method is applied to the inelastic Boltzmann equation. The  IMM is introduced in Sec.\ \ref{sec2bis} and the existing results in the literature for the zeroth (Euler) and first (NS) orders  in gradients are generalized to arbitrary $\qq$. In Sec.\ \ref{sec3}, the Burnett transport coefficient of the pressure tensor and the heat flux are calculated. They are explicitly given in terms of the coefficient of normal restitution, the dimensionality of the system, and the  parameter $\qq$. The most technical details are relegated to Appendixes \ref{appA} and \ref{appB}. The results are widely discussed in Sec.\ \ref{sec4}, where  the $\alpha$ dependence of the Burnett transport coefficients is presented and explicit expressions of the above coefficients in the elastic limit ($\alpha=1$) are given and compared with those in the literature. In addition, based on the formal structure of the results for the IMM,  estimates of the Burnett coefficients for the IHSM are displayed. Finally, the paper is closed in Sec.\ \ref{sec5} with some concluding remarks.

\section{From kinetic to hydrodynamic descriptions}
\label{sec2}

In this section, the CE method is described for a $d$-dimensional system composed by inelastic  particles of mass $m$ and coefficient of normal restitution $\alpha$ ($0\le \alpha \le 1$).

First, the Boltzmann equation for a force-free $d$-dimensional granular gas is considered,
\begin{equation}
  \label{eq:1}
  \left(\partial_t +\mn{v}\cdot \boldsymbol{\nabla}\right)f(\mn{r},\mn{v},t)=J[\mn{v}|f,f],
\end{equation}
where $f(\mn{r},\mn{v},t)$ is the distribution function of a particle at position $\mn{r}$, with velocity $\mn{v}$ at time $t$. The explicit form of the collision operator $J[\mn{v}|f,f]$ is so far not needed, except that it must reflect the collision rules relating the precollisional velocities $\{\mn{v}'_1,\mn{v}_2'\}$  to the postcollisional velocities $\{\mn{v}_1,\mn{v}_2\}$:
\begin{subequations}
\label{eq:4}
\beq
  \mn{v}_1'=\mn{v}_1-\frac{1}{2}(1+\alpha^{-1})(\mn{g}\cdot \hnn{\sigma})\hnn{\sigma},
\eeq
\beq
  \mn{v}_2'=\mn{v}_2+\frac{1}{2}(1+\alpha^{-1})(\mn{g}\cdot \hnn{\sigma})\hnn{\sigma},
\eeq
\end{subequations}
where $\mn{g}=\mn{v}_1-\mn{v}_2$ is the relative velocity and $\hnn{\sigma}$ is a unit vector directed along the line of centers of the two colliding particles.

Secondly, as usual, the  hydrodynamic fields are chosen to be the number density
\begin{equation}
  \label{eq:3}
  n(\mn{r},t)=\int d\mn{v}f(\mn{r},\mn{v},t),
\end{equation}
the flow velocity
\begin{equation}
  \label{eq:7}
  \mn{u}=\frac{1}{n}\int d\mn{v}\ \mn{v} f(\mn{v},\mn{r},t),
\end{equation}
and the granular temperature
\begin{equation}
  \label{eq:6}
  T(\mn{r},t)=\frac{m}{nd}\int d\mn{v}\ V^2 f(\mn{v},\mn{r},t),
\end{equation}
where $\mn{V}=\mn{v}-\mn{u}$ is the peculiar velocity.
By taking moments in the Boltzmann equation with respect to $1,\mn{v}$, and $v^2$, the balance equations are obtained:
\beq
  \label{eq:8}
\partial_t n+\boldsymbol{\nabla}\cdot (n\mn{u})=0,
\eeq
\beq
\partial_t u_i+u_j\nabla_j u_i+\frac{1}{\rho}\nabla_jP_{ij}=0,
\label{eq:8b}
\eeq
\beq
\partial_t T+\mn{u}\cdot \boldsymbol{\nabla} T+\frac{2}{nd}\left(\mathsf{P}:\boldsymbol{\nabla} \mn{u}+\boldsymbol{\nabla}\cdot \mn{q}\right)=-\zeta T.
 \label{eq:8c}
\eeq
In Eq.\ \eqref{eq:8b}, $\rho=mn$ is the mass density. The pressure tensor $\mathsf{P}$, the heat flux $\mn{q}$, and the cooling rate $\zeta$ are defined in terms of the distribution function as
\beq
  \label{eq:9}
\mathsf{P}(\mn{r},t)=m\int d\mn{v}\ \mn{V}\mn{V} f(\mn{r},\mn{v},t),
\eeq
\beq
  \label{eq:9a}
  \mn{q}(\mn{r},t)=\frac{m}{2} \int d\mn{v}\ {V}^2\mn{V} f(\mn{r},\mn{v},t),
\eeq
\beq
  \label{eq:9b}
 \zeta(\mn{r},t)=-\frac{m}{nTd}\int d\mn{v}\ v^2 J[\mn{v}|f,f].
\eeq
As the collision operator $J$ conserves the number of particles and linear momentum, in the expression of the cooling rate \eqref{eq:9b}, $v^2$ can be replaced by $V^2$.

Finally, the CE method is applied. This method provides a \emph{normal} solution to the Boltzmann equation, i.e., a solution where all space and time dependence occurs through the hydrodynamic fields,
\begin{equation}
  \label{eq:10}
  f(\mn{r},\mn{v},t)=f[\mn{v}|n(\mn{r},t),\mn{u}(\mn{r},t),T(\mn{r},t)],
\end{equation}
and, as a consequence, a closed hydrodynamic description is obtained. The functional dependence on the hydrodynamic fields in Eq.\ \eqref{eq:10} is made local in space by an expansion in spatial gradients as
\beq
  \label{eq:11}
  f(\mn{r},\mn{v},t)=f^{(0)}(\mn{r},\mn{v},t)+\epsilon f^{(1)}(\mn{r},\mn{v},t)  +\epsilon^2 f^{(2)}(\mn{r},\mn{v},t)+\cdots,
\eeq
where the superscript denotes the order of the gradient and $\epsilon$ is a non-uniformity parameter. In this way, the perturbative orders denoted by $\epsilon$ are associated with the gradients, and hence the hydrodynamic fields are of zeroth order. The mean difference of the CE method with respect to other perturbative schemes is the association of different time scales to different orders in gradients \cite{RL77}. Therefore, the time derivative is also expanded as
\begin{equation}
  \label{eq:12}
  \partial_t=\partial_t^{(0)}+\epsilon \partial_t^{(1)}+\epsilon^2 \partial_t^{(2)}+ \cdots .
\end{equation}

Once the ingredients of the method have been put together, the distribution function (and also a closed set of hydrodynamic equations)  are obtained at the desired order in the gradients. In particular,  the pressure tensor, the heat flux, and the cooling rate can be written as
\beq
  \label{eq:13}
\mathsf{P}=\mathsf{P}^{(0)}+\epsilon \mathsf{P}^{(1)}+\epsilon^2 \mathsf{P}^{(2)}+\cdots,
\eeq
\beq
  \mn{q}=\mn{q}^{(0)}+\epsilon \mn{q}^{(1)}+\epsilon^2 \mn{q}^{(2)}+\cdots,
\eeq
\beq
\label{eq:13c}
 \zeta=\zeta^{(0)}+\epsilon \zeta^{(1)}+\epsilon^2 \zeta^{(2)}+\cdots,
\eeq
where the different powers of $\epsilon$ correspond to retaining the orders of the expansion \eqref{eq:11} of the distribution function in the definitions \eqref{eq:9}--\eqref{eq:9b}.
When the zeroth and first-order terms in Eqs.\ \eqref{eq:13}--\eqref{eq:13c} are inserted into the balance equations \eqref{eq:8}--\eqref{eq:8c}, the Euler and NS hydrodynamic equations are obtained, respectively. The second-order terms yield the Burnett hydrodynamic equations. As said in Sec.\ \ref{sec1}, the main objective of this paper is to derive the Burnett constitutive equations for the IMM with explicit expressions for all the involved transport coefficients.

\section{Inelastic Maxwell models. Euler and Navier--Stokes orders}
\label{sec2bis}
The IMM collisional {operator is} \cite{GS11}
\begin{align}
  \label{eq:2}
J[\mn{v}_1|f,f]=&\frac{(d+2)\nu_0}{2n\Omega_d}\int d\mn{v}_2\int d\hnn{\sigma} \left[\alpha^{-1} f(\mn{r},\mn{v}_1',t) \right. \nonumber \\
       &\left. \times f(\mn{r},\mn{v}_2',t)-f(\mn{r},\mn{v}_1,t)f(\mn{r},\mn{v}_2,t)\right],
\end{align}
where $\Omega_d=2\pi^{d/2}/\Gamma(d/2)$ is the total solid angle in $d$ dimensions and $\nu_0$ is an effective collision frequency that is taken here to be proportional to the density and the $\qq$ power of the temperature,
\begin{equation}
  \label{eq:5}
    \nu_0\propto nT^{\qq}.
\end{equation}
The factor $(d+2)/2$ appearing on the right-hand side of Eq.\ \eqref{eq:2} has been introduced to guarantee that the NS shear viscosity in the elastic limit ($\alpha=1$) is simply $\eta_0=p/\nu_0$, where $p=nT$ is the hydrostatic pressure.
The class of models with  general $\qq$ {mimic} other inelastic models with a collision rate proportional to a power of the relative velocity \cite{ETB06a,TBE07,Y13}.

The specific IMM form \eqref{eq:2} allows one to exactly express any collisional moment of degree $k$ in terms of the moments  of $f$ of degree equal to or smaller than $k$. In particular, the cooling rate $\zeta$ is \cite{S03}
\begin{equation}
  \label{eq:18}
  \zeta=\frac{d+2}{4d}(1-\alpha^2)\nu_0.
\end{equation}
As a consequence, $\zeta$ does not depend on the hydrodynamic gradients and hence Eq.\ \eqref{eq:13c} implies
\beq
  \label{eq:18a}
  \zeta=\zeta^{(0)},
  \eeq
  \beq
  \zeta^{(i)}=0, \quad i\ge 1.
\eeq
In the case of the IHSM, $\zeta^{(1)}=0$ in the dilute limit \cite{BDKS98,GD99} but $\zeta^{(2)}\neq 0$, although its influence on the energy balance equation is relatively very small.

\subsection{Euler order}
To  zeroth order, Eq.\ \eqref{eq:1} becomes,
\begin{equation}
  \label{eq:14}
  \partial_t^{(0)}f^{(0)}=J[\mn{v}|f^{(0)},f^{(0)}].
\end{equation}
As the time dependence of $f^{(0)}$ occurs through the hydrodynamic fields, the time derivative can be written as
\begin{equation}
  \label{eq:15}
  \partial_t^{(0)}f^{(0)}=\frac{\partial f^{(0)}}{\partial n}\partial_t^{(0)}n
+\frac{\partial f^{(0)}}{\partial u_i}\partial_t^{(0)}u_i+\frac{\partial f^{(0)}}{\partial T}\partial_t^{(0)}T.
\end{equation}
The balance equations \eqref{eq:8}--\eqref{eq:8c} to zeroth order read $\partial^{(0)}_tn=\partial^{(0)}_tu_i=0$ and
  \beq
  \label{eq:16c}
  \partial^{(0)}_tT=-\zeta T,
\eeq
where in Eq.\ \eqref{eq:16c} we have taken into account Eq.\ \eqref{eq:18a}.
Using Eq.\ \eqref{eq:16c}, Eq.\ \eqref{eq:14} becomes
\begin{equation}
  \label{eq:17}
  -T\zeta\partial_Tf^{(0)}=J[\mn{v}|f^{(0)},f^{(0)}].
\end{equation}
As Eq.\ \eqref{eq:17} is also verified by the distribution function of the homogeneous cooling state (HCS), the distribution function of zeroth order $f^{(0)}$ is the local version of the latter with the replacement $\mn{v}\to\mn{V}$. Since $f^{(0)}$ is an isotropic function (with respect to $\mn{V}$), then
\beq
  \label{eq:21}
  P_{ij}^{(0)}=p\delta_{ij}=nT\delta_{ij},
  \eeq
  \beq
   \mn{q}^{(0)}=0.
   \eeq

Although the solution to Eq.\ \eqref{eq:17} is not known, its velocity moments can be in principle obtained in a recursive way \cite{BK03,EB02b}. In particular, the fourth-degree cumulant is \cite{S03}
\begin{equation}
  \label{eq:22}
  \cum\equiv\frac{d}{d+2}\frac{\medio{V^4}}{\medio{V^2}^2}-1=\frac{6(1-\alpha)^2}{4d-7+3\alpha(2-\alpha)},
\end{equation}
where $\langle A(\mn{V}) \rangle=n^{-1} \int d\mn{v}  A(\mn{V})  f^{(0)}$.

\subsection{NS order}
\label{sec3b}
Once $f^{(0)}$ is characterized, it is possible to consider the first order. Now, the equation for $f^{(1)}$ reads
\beq
  \label{eq:23}
  {\partial_t^{(0)}f^{(1)}+\mathcal{L}f^{(1)}=-\left(\partial_t^{(1)}+\mn{v}\cdot \boldsymbol{\nabla}\right)f^{(0)},}
\eeq
where
\beq
{\mathcal{L}f^{(1)}=-J^{(1)}[f,f]=-J[f^{(1)},f^{(0)}] -J[f^{(0)},f^{(1)}]}
\eeq
{is the linearized (inelastic) Boltzmann collision operator acting on $f^{(1)}$.}

The right-hand side  of Eq.\ \eqref{eq:23} can  easily be evaluated taking into account that the balance equations to first order become
\beq
  \label{eq:24}
  \partial_t^{(1)} n+\boldsymbol{\nabla}\cdot (n\mn{u})=0,
  \eeq
  \beq
  \label{eq:24a}
  \partial_t^{(1)} u_i+u_j\nabla_j u_i+{\rho}^{-1}\nabla_i p=0,
  \eeq
  \beq
  \label{eq:24b}
   \partial_t^{(1)} T+\mn{u}\cdot \boldsymbol{\nabla} T+\frac{2}{d}T\boldsymbol{\nabla}\cdot \mn{u}=0.
\eeq

{Equation \eqref{eq:23} is a linear integral equation for $f^{(1)}$ with an inhomogeneous term given by the right-hand side. It is straightforward to check that the inhomogeneous term is orthogonal to $(1,\mathbf{v}, v^2)$, i.e., the subspace
associated with the null eigenvalue of the linear operator acting on $f^{(1)}$ (solubility conditions) \cite{FK72}.}
The general solution to Eq.\ \eqref{eq:23} is of the  form \cite{BDKS98,S03},
\begin{align}
  \label{eq:25}
     f^{(1)}(\mn{r},\mn{v},t)=&\mathcal{A}_i(\mn{V}) \nabla_i \ln n+\mathcal{{B}}_i(\mn{V}) \nabla_i \ln T
      \nn
      &+ \mathcal{C}_{ij}(\mn{V})\left(\nabla_i u_j+\nabla_j u_i-\frac{2}{d}\delta_{ij}\boldsymbol{\nabla}\cdot\mn{u}\right),\nn
\end{align}
where $\mathcal{A}_i(\mn{V})$, $\mathcal{B}_i(\mn{V})$, and $\mathcal{C}_{ij}(\mn{V})$ obey a set of linear integral equations. The absence of an independent scalar term proportional to $\boldsymbol{\nabla}\cdot \mn{u}$ in Eq.\ \eqref{eq:25} implies that any isotropic moment of $f^{(1)}$ must vanish. On the other hand, for dense gases the above property does not apply \cite{GS95,GD99}.

The NS constitutive equations for the pressure tensor and the heat flux have the form,
\beq
  \label{eq:26}
  P_{ij}^{(1)}=-\eta\left(\nabla_iu_j+\nabla_ju_i-\frac{2}{d}\delta_{ij}\boldsymbol{\nabla}\cdot \mn{u}\right),
  \eeq
  \beq
  \label{eq:26a}
  \mn{q}^{(1)}=-\mu\boldsymbol{\nabla} n-\kappa\boldsymbol{\nabla} T,
  \eeq
where, by dimensional analysis,  the shear viscosity $\eta$, the thermal conductivity $\kappa$, and the coefficient $\mu$ have the following scaling properties:
 \beq
 \eta=\eta_0 \eta^*(\alpha), \quad \kappa=\kappa_0 \kappa^*(\alpha), \quad \mu=\frac{T\kappa_0}{n}\mu^*(\alpha).
  \eeq
Here, $\eta_0={p}/{\nu_0}$ and  $\kappa_0=[{d(d+2)}/{2(d-1)}]{\eta_0}/{m}$ are the shear viscosity and thermal conductivity coefficients in the elastic limit.

In the case of the IMM, the transport coefficients can be obtained without the need of determining the unknown functions $\mathcal{A}_i(\mn{V})$, $\mathcal{B}_i(\mn{V})$, and $\mathcal{C}_{ij}(\mn{V})$. The method consists of multiplying Eq.\ \eqref{eq:23} by $V_iV_j-d^{-1}V^2\delta_{ij}$ and $V^2 \mn{V}$, integrating over velocity, and applying Eqs.\ \eqref{eq:24}--\eqref{eq:24b}. The results are
\beq
  \label{eq:28}
\eta^*=\frac{1}{\nu_{0|2}^*-(1-\qq)\zeta^*},
\eeq
\beq
\label{eq:28b}
  \kappa^*=\frac{d-1}{d}\frac{1+2\cum}{\nu_{2|1}^*-2\zeta^*},
  \eeq
  \beq
  \label{eq:28c}
  \mu^*=\frac{\kappa^*}{1+2\cum}\frac{\zeta^*+
    \nu_{2|1}^*\cum}{\nu_{2|1}^*-(2-\qq)\zeta^*}.
\eeq
Upon deriving these equations, use has been made of the exact expressions for the second- and third-degree collisional moments for IMM [see Eqs.\ (2.17) and (2.20) of Ref.\ \cite{GS07}]. In Eqs.\ \eqref{eq:28}--\eqref{eq:28c},
\begin{equation}
  \label{eq:a1_6}
   \nu_{0|2}^*=\frac{(1+\alpha)(d+1-\alpha)}{2d},
   \eeq
       \beq
      \label{eq:a1_6c}
       \nu_{2|1}^*=\frac{(1+\alpha)\left[5d+4-\alpha(d+8)\right]}{8d},
       \eeq
and $\zeta^*=\zeta/\nu_0$.
Equations \eqref{eq:28}--\eqref{eq:28c} for $\qq=\frac{1}{2}$ were first obtained in Ref.\ \cite{S03}. Here, they are generalized to arbitrary $\qq$.

It is interesting to observe that the structure of Eqs.\ \eqref{eq:28}--\eqref{eq:28c} for the NS transport coefficients (with $\qq=\frac{1}{2}$) coincides with that of the IHSM, except that the $\alpha$ dependence of the cumulant $c$, the cooling rate $\zeta^*$, and the collision frequencies $\nu_{0|2}^*$ and $\nu_{2|1}^*$ are different \cite{BDKS98,BC01,vNE98,MS00,GSM07}. The IHSM expressions can be found in Appendix \ref{appD}.

\section{Burnett order}
\label{sec3}
In this section, the Burnett constitutive equations  for the pressure tensor and heat flux are derived and the corresponding transport coefficients are evaluated.  The procedure is similar to the one followed at zeroth and first orders and makes use of the preceding results.
As is usually done in the case of elastic collisions \cite{CC70,FK72,M89}, we will choose $p=nT$ instead of $n$ as a hydrodynamic variable in the Burnett order.

To second order in $\epsilon$, the kinetic equation for $f$ reads
\beq
  \label{eq:30}
  \partial_t^{(0)} f^{(2)}+\left(\partial_t^{(1)}+\mn{v}\cdot
\boldsymbol{\nabla} \right)f^{(1)}+\partial_t^{(2)} f^{(0)} =J^{(2)}[f,f],
\eeq
where
\beq
J^{(2)}[f,f]=J[f^{(2)},f^{(0)}]+J[f^{(0)},f^{(2)}]+J[f^{(1)},f^{(1)}].
\eeq
{Equation \eqref{eq:30} can be rewritten as
\bal
  \label{eq:30bis}
  \partial_t^{(0)} f^{(2)}+\mathcal{L}f^{(2)}=&-\left(\partial_t^{(1)}+\mn{v}\cdot
\boldsymbol{\nabla} \right)f^{(1)}-\partial_t^{(2)} f^{(0)}\nn
& +J[f^{(1)},f^{(1)}].
\eal
As in the case of Eq.\ \eqref{eq:23}, the inhomogeneous term (right-hand side) of Eq.\ \eqref{eq:30bis} is orthogonal to $(1,\mathbf{v},v^2)$, so that the solubility conditions are satisfied.}
 To evaluate $\partial_t^{(2)}f^{(0)}$, the balance equations to second order are needed,
\beq
  \label{eq:31}
   \partial_t^{(2)} n=0,
   \eeq
   \beq
  \label{eq:31a}
  \partial_t^{(2)} u_i+\frac{1}{\rho}\nabla_jP_{ij}^{(1)}=0,
  \eeq
  \beq
  \label{eq:31b}
   \partial_t^{(2)} T+\frac{2}{nd}\left(P_{ij}^{(1)}\nabla_i \mn{u}_j+\boldsymbol{\nabla}\cdot \mn{q}^{(1)}\right)=0.
\eeq

The aim of this section is to determine the pressure tensor and heat flux to second order in the spatial gradients. This is accomplished by taking the corresponding moments in Eq.\ \eqref{eq:30}. Each quantity will be considered separately.
Since the algebra involved is rather cumbersome, we give here the final results, the mathematical details being postponed to Appendixes \ref{appA} and \ref{appB}.

\subsection{Pressure tensor}
The Burnett constitutive equation for the pressure tensor
$P_{ij}^{(2)}$ can be written as
\begin{widetext}
\bal
\label{e33}
P_{ij}^{(2)}=&a_1
\frac{\kappa_0}{\nu_0}\left(\nabla_i\nabla_jT-
\frac{1}{d}\delta_{ij}\nabla^2T\right)+a_2\frac{{T}\kappa_0}{p\nu_0}\left(\nabla_i\nabla_jp-
\frac{1}{d}\delta_{ij}\nabla^2p\right)+a_3
\frac{\kappa_0}{T\nu_0}\left[(\nabla_iT)(\nabla_jT)-
\frac{1}{d}\delta_{ij}(\boldsymbol{\nabla} T)^2\right]\nn
&+a_4\frac{{T}\kappa_0}{p^2\nu_0}\left[(\nabla_ip)(\nabla_jp)-
\frac{1}{d}\delta_{ij}(\boldsymbol{\nabla} p)^2\right]+a_5\frac{\kappa_0}{p\nu_0}
\left[(\nabla_iT)(\nabla_jp)+(\nabla_ip)(\nabla_jT)-\frac{2}{d}\delta_{ij}(\boldsymbol{\nabla}
p) \cdot (\boldsymbol{\nabla} T)\right]  \nonumber \\
&+a_6
\frac{\eta_0}{\nu_0}D\left(D_{ij}-\frac{1}{d}\delta_{ij}D\right) +a_7 \frac{\eta_0}{\nu_0} \Big[D_{ik}D_{kj}-\omega_{ik}\omega_{kj}
-\frac{1}{d}\delta_{ij} \left(D_{\ell k}D_{k\ell}-\omega_{\ell k}\omega_{k\ell}\right)+\omega_{ik}D_{kj}- D_{ik}\omega_{kj}\Big],
\eal
\end{widetext}
where
\beq
\label{e32}
 D\equiv\boldsymbol{\nabla} \cdot \mn{ u},
 \eeq
 \beq
 D_{ij}\equiv\frac{1}{2}\left(\nabla_i u_j+\nabla_j u_i\right),
 \eeq
 \beq
 \omega_{ij}\equiv\frac{1}{2}\left(\nabla_j u_i-\nabla_i u_j\right).
\eeq
The terms in Eq.\ \eqref{e33} fall into two classes \cite{M89}: those which are linear in second derivatives of $T$ and $p$  and those which are quadratic in the first derivatives of $T$, $p$, and $\mathbf{u}$. The coefficients $a_1$ and $a_2$ correspond to the first class, while the coefficients $a_3$--$a_7$ correspond to the second class.

The reduced Burnett coefficients $a_i$ are dimensionless quantities that are  consistently determined  in Appendix \ref{appA}.
While the coefficients $a_1$--$a_5$ (involving terms associated with pressure and temperature gradients) obey a set of coupled linear equations, the coefficients $a_6$ and $a_7$ are decoupled from the rest. They are given by
\begin{equation}
\label{e40}
a_6=\frac{2}{d}\frac{d-2(2-\qq)}{\nu_{0|2}^*-(1-2\qq)\zeta^*}\eta^*,
\end{equation}
\begin{equation}
\label{e41}
a_7=\frac{2\eta^*}{\nu_{0|2}^*-(1-2\qq)\zeta^*}.
\end{equation}
{As shown in Appendix \ref{nonNewt}, the coefficients $a_6$ and $a_7$ agree with the results obtained in the zero strain rate limit of the viscometric functions defined in the non-Newtonian uniform shear and uniform longitudinal flows.}

The two  linear Burnett coefficients $a_1$ and $a_2$ obey a closed set of two equations whose solution is
\bal
a_1=&\frac{4}{(d+2)\Delta}\Big\{\left[\nu_{0|2}^*-(3-2\qq)\zeta^*\right]\left(\kappa^*-\mu^*\right)\nn
&+{(1-\qq)\zeta^*}
\left({\frac{d-1}{d}}\eta^*-\mu^*\right)\Big\},
\label{a2}
\eal
\beq
a_2=-\frac{4}{(d+2)\Delta}\left[
{\frac{d-1}{d}}-\frac{\mu^*}{\eta^*}-\zeta^*\left(\kappa^*-\mu^*\right)\right],
\label{a4}
\eeq
where
\beq
\Delta\equiv \left[\nu_{0|2}^*-(2-\qq)\zeta^*\right]\left[\nu_{0|2}^*-2(1-\qq)\zeta^*\right].
\label{Delta}
\eeq
The remaining three  coefficients are given by
\beq
\left(\begin{array}{c}
  a_3\\a_4\\a_5
\end{array}
\right)=\mathsf{L}^{-1}\cdot \mathsf{X},
\label{e42}
\eeq
where $\mathsf{L}$ is the square matrix,
\begin{equation}
\label{e43.1}
\mathsf{L}=\left(
\begin{array} {ccc}
\nu_{0|2}^*&0&{2}(1-\qq)\zeta^*\\
0&\nu_{0|2}^*-2(2-\qq)\zeta^*&-{2}\zeta^*\\
-\zeta^*&(1-\qq)\zeta^*&\nu_{0|2}^*-(2-\qq)\zeta^*
\end{array}
\right),
\end{equation}
and $\mathsf{X}$ is the column matrix
\bal
\label{e44}
\mathsf{X}=&\frac{4}{d+2}
\left(
\begin{array}{c}
(1-\qq)(\kappa^*-\mu^*)\\
{\frac{d-1}{d}}\eta^*-\mu^*\\
(1-\frac{\qq}{2})\mu^*-{\frac{d-1}{2d}}\eta^*
\end{array}
\right)\nn
&-\left(\begin{array}{cc}
\qq(1-\qq)\zeta^*&-(1-\qq)(2-\qq)\zeta^*\\
0&-2\zeta^*\\
-\qq\zeta^*&2(1-\qq)\zeta^*
\end{array}
\right)\cdot
\left(\begin{array}{c}
a_1\\
a_2
\end{array}
\right).
\eal

\subsection{Heat flux}
The structure of the Burnett constitutive equation for the heat flux $\mn{q}^{(2)}$ is
\bal
\label{e56}
q_i^{(2)}=&b_1\frac{T\kappa_0}{\nu_0}\nabla^2 u_i+b_2\frac{T\kappa_0}{\nu_0}\nabla_iD
+b_3\frac{\kappa_0}{\nu_0}D_{ij}\nabla_jT\nn
&+b_4\frac{\eta_0}{\rho\nu_0}D_{ij}\nabla_jp
+b_5\frac{\kappa_0}{\nu_0}\omega_{ij}\nabla_jT+{b_6 \frac{\eta_0}{\rho\nu_0} \omega_{ij}\nabla_jp}\nn
&+b_7\frac{\kappa_0}{\nu_0}D\nabla_iT+b_8\frac{\eta_0}{\rho\nu_0}D\nabla_ip.
\eal
Analogously to the case of the pressure tensor, Eq.\ \eqref{e56} contains linear Burnett terms (with coefficients $b_1$ and $b_2$) and nonlinear Burnett terms (with coefficients $b_3$--$b_8$).

The procedure to obtain the coefficients $b_i$ is described in Appendix \ref{appB}. In the case of the linear Burnett coefficients $b_1$ and $b_2$, the results are
\begin{equation}
\label{e62}
b_1=\frac{d-1}{d(d+2)}\frac{\psi-(d+2)\eta^*}{\nu_{2|1}^*-2(1-\qq)\zeta^*},
\end{equation}
\begin{equation}
\label{e61}
b_2={\frac{\frac{(d-1)(d-2)}{d^2(d+2)} \left[\psi-(d+2)\eta^*\right]-\frac{2}{d}\kappa^*-\mu^*}{\nu_{2|1}^*-2(1-\qq)\zeta^*}},
\end{equation}
where
\begin{equation}
  \label{eq:62a}
  \psi\equiv\frac{(d+4)(1+\cum)+d \lambda^*\eta^*}{\nu_{2|2}^*-(2-\qq)\zeta^*},
\end{equation}
with
\bal
\nu_{2|2}^*=&\frac{(1+\alpha)}{8d(d+4)} \left[7d^2+31d+18 -\alpha(d^2+14d+34)\right. \nn
  & \left. +3\alpha^2(d+2)-6\alpha^3\right],
 \label{eq:62c}
\eal
  \begin{equation}
  \label{eq:a1_7}
  \lambda^*=\frac{(1+\alpha)^2}{8d^2}\left[d^2+5d-2-3\alpha(d+4)+6\alpha^2\right].
\end{equation}

The remaining coefficients obey pairs of linear equations whose solutions are
\beq
\label{b3b4}
\left(\begin{array}{c}
  b_3\\b_4
\end{array}
\right)=\mathsf{M}^{-1}\cdot \left(\begin{array}{c}
  A_1\\A_2
\end{array}
\right),
\eeq
\beq
\left(\begin{array}{c}
  b_5\\b_6
\end{array}
\right)=\mathsf{M}^{-1}\cdot \left(\begin{array}{c}
  B_1\\B_2
\end{array}
\right),
\eeq
\beq
\label{b7b8}
\left(\begin{array}{c}
  b_7\\b_8
\end{array}
\right)=\mathsf{M}^{-1}\cdot \left(\begin{array}{c}
  C_1\\C_2
\end{array}
\right),
\eeq
where
\beq
\mathsf{M}=\left(
\begin{array}{cc}
  \nu_{2|1}^*-(1-\qq)\zeta^*&-\frac{2(d-1)}{d(d+2)}{(\qq-1)}\zeta^*\\
 {-}\frac{d(d+2)}{2(d-1)}\zeta^*&\nu_{2|1}^*-(3-2\qq)\zeta^*
 \end{array}
 \right),
 \eeq
\beq
A_1={\frac{2(d-1)}{d}}\left[\frac{2-\qq}{d+2}\psi
-(1-\qq)\eta^*\right]
+\frac{4(\kappa^*-\mu^*)}{d+2},
\eeq
\beq
A_2=-2\left(\eta^*-\frac{d}{d-1}\mu^*\right),
\eeq
\beq
B_1={2(\kappa^*-{\mu^*})},\quad B_2={\frac{d+2}{d-1}\mu^*},
\eeq
\bal
C_1=&-{ \frac{2(d-1)}{d^2}}\left[\frac{2-\qq}{d+2}\psi -(1-\qq)\eta^*\right] \nonumber\\
& +\frac{d^2+2\qq(d+2)-8}{d(d+2)} \left(\kappa^*-\mu^*\right),
\eal
\beq
C_2=\frac{2}{d}\eta^*+\frac{{d^2+2\qq(d+2)-8}}{2(d-1)}\mu^*.
\label{CC2}
\eeq

\section{Discussion}
\label{sec4}
\subsection{Structure of the Burnett coefficients}

The main results of the paper, derived for the IMM, are summarized by Eqs.\ \eqref{e33} and \eqref{e56}, complemented by the explicit expressions for the (reduced) Burnett transport coefficients $\{a_i\}$ and $\{b_i\}$. They are exactly given by Eqs.\ \eqref{e40}--\eqref{e42}, \eqref{e62}, \eqref{e61}, and \eqref{b3b4}--\eqref{b7b8} for arbitrary values of the dimensionality $d$, the model parameter $\qq$, and the coefficient of normal restitution $\alpha$.

In principle, the coefficients of $D_{ik}D_{kj}-\frac{1}{d}\delta_{ij} D_{\ell k}D_{k\ell}$, $\frac{1}{d}\delta_{ij} \omega_{\ell k}\omega_{k\ell}-\omega_{ik}\omega_{kj}$, and $\omega_{ik}D_{kj}- D_{ik}\omega_{kj}$ in Eq.\ \eqref{e33} do not need to be the same. However, our results show that the three coefficients degenerate into a single one ($a_7$) in the IMM.

The seven coefficients $\{a_i\}$ associated with the pressure tensor depend on $\alpha$ \emph{only} through a dependence on the reduced cooling rate $\zeta^*$ and the three dimensionless NS coefficients $\eta^*$, $\kappa^*$, and $\mu^*$. Note that the dependence on $\nu_{0|2}^*$ can be eliminated in favor of $\eta^*$ and $\zeta^*$ via Eq.\ \eqref{eq:28}. Therefore, there must exist only three \emph{$\alpha$-independent} equations relating  the seven coefficients $\{a_i\}$. One of those relations is, simply,
\beq
\frac{a_6}{a_7}=1-\frac{2(2-\qq)}{d}.
\label{a6a7}
\eeq

In the case of the eight coefficients $\{b_i\}$ associated with the heat flux, they depend on $\alpha$ through the same coefficients as before ($\zeta^*$, $\eta^*$, $\kappa^*$, and $\mu^*$) plus the coefficient $\psi$. Note that the dependence on $\nu_{2|1}^*$ can be eliminated in favor of $\kappa^*$, $\mu^*$, and $\zeta^*$ via Eqs.\ \eqref{eq:28b} and \eqref{eq:28c}. Therefore, there must exist again only three independent relations among the  coefficients $\{b_i\}$.
Finally, the 15 coefficients $\{a_i\}$ and $\{b_i\}$ depend on the five coefficients $\zeta^*$, $\eta^*$, $\kappa^*$,  $\mu^*$, and $\psi$, so that the total number of constraints is 10. Since six of them involve either only the $\{a_i\}$ or the $\{b_i\}$, there are four conditions relating all the coefficients.

The Burnett constitutive equations \eqref{e33} and \eqref{e56} can be written in other equivalent forms \cite{CC70,FK72,M89}. In particular, in the form found in Chapman and Cowling's  standard textbook \cite{CC70}, the pressure tensor is
\begin{widetext}
 \bal
P_{ij}^{(2)}=&\varpi_1 \frac{\eta_0^2}{p}D
\widetilde{D}_{ij}-\frac{\eta_0^2}{2p}\Delta_{ijk\ell}\Big\{\varpi_2\rho^{-1}\nabla_k\nabla_\ell p+\varpi_2'(\nabla_k\rho^{-1})(\nabla_\ell p)+\varpi_2''
\left[(\nabla_k u_m)(\nabla_m
u_\ell)+2(\nabla_k u_m)\widetilde{D}_{m\ell}\right]\nn
&-\varpi_3 \frac{p}{\rho
T}\nabla_k\nabla_\ell T-\varpi_4 \frac{1}{\rho
T}(\nabla_k p)(\nabla_\ell T)-\varpi_5 \frac{p}{\rho T^2}(\nabla_k
T)(\nabla_\ell T)-\varpi_6 \widetilde{D}_{km}\widetilde{D}_{m\ell}\Big\},
\label{CCP}
\eal
\end{widetext}
where
\beq
\widetilde{D}_{ij}\equiv D_{ij}-\frac{D}{d}\delta_{ij},\quad
\Delta_{ijk\ell}\equiv
\delta_{ik}\delta_{j\ell}+\delta_{i\ell}\delta_{jk}-\frac{2}{d}\delta_{ij}\delta_{k\ell}.
\eeq
The  first column of Table \ref{table1} shows the relations between the coefficients $\varpi_i$ of Eq.\ \eqref{CCP} and the coefficients $a_i$ of Eq.\ \eqref{e33}.
Note that in Ref.\ \cite{CC70}, which is restricted to elastic gases, $\varpi_2=\varpi_2'=\varpi_2''$. This degeneracy is broken in the inelastic case.
It is also interesting to note that, according to Eq.\ \eqref{a6a7}, the following relations hold for any $\alpha$:
\beq
\varpi_6=4\varpi_2''=\frac{4d}{d+2+2\qq}\varpi_1.
\eeq

\begin{table}[t]
\caption{Relationship between the coefficients  $\varpi_i$ and $a_i$ and between the coefficients  $\theta_i$ and $b_i$.
\label{table1}}
\begin{ruledtabular}
\begin{tabular} {lc}
$P_{ij}^{(2)}$ coefficients&$q_i^{(2)}$ coefficients\\
\hline
$\varpi_1=a_6+\frac{6}{d}a_7$&$\theta_1=\frac{d(d+2)}{2(d-1)}\left(\frac{d-2}{d}b_1-b_2+\frac{b_3+b_5}{d}+b_7\right)$\\
$\varpi_2=-\frac{d(d+2)}{2(d-1)}a_2$& $\theta_2=\frac{d^2(d+2)}{4(d-1)}\left(\frac{d-2}{d}b_1-b_2\right)$\\
$\varpi_2'=\frac{d(d+2)}{2(d-1)}a_4$&$ \theta_2'=\frac{d(d+2)}{4(d-1)}b_5$\\
$\varpi_2''=a_7$&$\theta_3=b_4+b_6$ \\
$\varpi_3=\frac{d(d+2)}{2(d-1)}a_1$&$\theta_4=\frac{d(d+2)}{d-1}b_1$ \\
$\varpi_4=\frac{d(d+2)}{2(d-1)}(2a_5+a_4)$& $\theta_5=\frac{d(d+2)}{6(d-1)}\left(b_3+b_5\right)$\\
$\varpi_5=\frac{d(d+2)}{2(d-1)}a_3$&$\theta_6=\frac{1}{d}\left(b_4+b_6\right)+b_8$ \\
$\varpi_6=4a_7$&$\theta_7=-b_6$ \\
\end{tabular}
\end{ruledtabular}
\end{table}

In the case of the heat flux, one can rewrite Eq.\ \eqref{e56} as \cite{CC70}
\bal
q_i^{(2)}=&\frac{\eta_0^2}{\rho T}\left[\theta_1 D\nabla_i
T-\frac{2}{d}\theta_2\nabla_i(DT)-2\theta_2'(\nabla_i u_j)(\nabla_j
T)\right.\nn
&+\theta_3\frac{T}{p}\widetilde{D}_{ij}\nabla_j p+\theta_4T\nabla_j
\widetilde{D}_{ij}+3\theta_5\widetilde{D}_{ij}\nabla_j T\nn
&\left.+\theta_6 \frac{T}{p}D\nabla_i p+\theta_7\frac{T}{p}(\nabla_i u_j)(\nabla_j p)\right],
\label{CCq}
\eal
where the relations between the  coefficients $\theta_i$ defined in Eq.\ \eqref{CCq} and $b_i$ defined in Eq.\ \eqref{e56} are given  in the second column of Table \ref{table1}.

\subsection{Elastic limit}

Before analyzing the $\alpha$ dependence of the Burnett coefficients, it is worthwhile considering the elastic limit ($\alpha=1$). In that case, one has $c=\zeta^*=\mu^*=0$, $\eta^*=\kappa^*=\nu_{0|2}^*=1$, $\nu_{2|1}^*=(d-1)/d$, and $\psi=d+4$. Inserting those values into the expressions  of the coefficients $a_i$ and $b_i$, one obtains the values displayed in the first column of Table \ref{table2}. The values of the alternative coefficients $\varpi_i$ and $\theta_i$ are then obtained from the expressions in Table \ref{table1}, the results being shown in the second column of Table \ref{table2}.

Obviously, some of the Burnett coefficients depend explicitly on the parameter $\qq$. In the elastic case, it is legitimate to relate that parameter with the power of a repulsive interaction potential $\phi(r)\sim r^{-s}$ as $s=2(d-1)/(1-2\qq)$. In that case, it is well known that $\eta_0\propto T^{1-\gamma}$. The generalization to \emph{any} (short-range) interaction potential can be simply achieved by the replacement $\gamma\to 1-\partial\ln \eta_0/\partial\ln T$ in Table \ref{table2}.
Particularizing to three-dimensional systems ($d=3$), one then recovers the expressions for the coefficients $\varpi_i$ and $\theta_i$ given in Ref.\ \cite{CC70} for an arbitrary potential in the first (Sonine) approximation.
This is a stringent consistency test of the results derived in this paper. Moreover, Table \ref{table2}, with the replacement $\gamma\to 1-\partial\ln \eta_0/\partial\ln T$, provides a generalization to any dimensionality of the Burnett coefficients given by Ref.\ \cite{CC70} in the first approximation. To the best of our knowledge, this generalization had not been derived before. {In particular, the results of Table \ref{table2} show that the exact universal relations \cite{CC70,TM80,S02} $\varpi_3=\theta_4$, $\varpi_3+\varpi_4+\theta_3=0$, $(d/2)\varpi_1/\varpi_2=(d/2)\theta_1/\theta_2=(d+4)/2-\partial\ln \eta_0/\partial\ln T$ hold for any dimensionality.}

It is instructive to note that the simple relations (holding in the elastic case) $\varpi_2=\varpi_2'=\varpi_2''$, $\varpi_4=0$, $\theta_2=\theta_2'$, and $\theta_6=\theta_7=0$ disappear if $\alpha<1$. Except for $\varpi_2=\varpi_2'$, this is due to the presence of the NS transport coefficient $\mu$ in the coefficients $c_{P,i}$ and $c_{q,i}$ appearing in Eqs.\ \eqref{e31} and \eqref{e55}. On the other hand, while $c_{P,2}=-pc_{P,4}$, the relation $a_2=-a_4$, and hence $\varpi_2=\varpi_2'$, is broken down in the inelastic case due to the fact that $\partial_t^{(0)}T\neq 0$.

\begin{table}[t]
\caption{Burnett coefficients in the elastic limit ($\alpha=1$).
\label{table2}}
\begin{ruledtabular}
\begin{tabular} {lc}
Eqs.\ \protect\eqref{e33} and  \protect\eqref{e56}&Eqs.\ \protect\eqref{CCP} and \protect\eqref{CCq}\\
\hline
$ a_1=\frac{4}{d+2}$&$\varpi_1= \frac{2(d+2+2\qq)}{d}$\\
$a_2=-\frac{4(d-1)}{d(d+2)}$&$\varpi_2=2$
\\
$ a_3=\frac{4(1-\qq)}{d+2}$&$\varpi_2'=\varpi_2''=2$\\
$a_4=\frac{4(d-1)}{d(d+2)}$&$\varpi_3= \frac{2d}{d-1}$\\
$a_5=-\frac{2(d-1)}{d(d+2)}$&$\varpi_4= 0$\\
$a_6=\frac{2(d-4+2\qq)}{d}$&$\varpi_5= \frac{2d(1-\qq)}{d-1}$\\
$a_7=2$&$\varpi_6= 8$\\
 $b_1=\frac{2}{d+2}$&${\theta_1= \frac{d(d+2)( d+2 + 2 \qq)}{2(d-1)^2}}$\\
 $b_2=-\frac{2(5d-2)}{d(d-1)(d+2)}$&$\theta_2= \frac{d^2(d+2)}{2(d-1)^2}$\\
 $b_3=\frac{2\left[d^2+7d-6-2(d-1)\qq\right]}{(d-1)(d+2)}$&$\theta_2'= \frac{d^2(d+2)}{2(d-1)^2}$\\
 $b_4=-\frac{2d}{d-1}$&$\theta_3= -\frac{2 d}{d-1}$\\
  $   b_5=\frac{2d}{d-1}$&$\theta_4= \frac{2 d}{d-1}$\\
   $  b_6=0$&$\theta_5= \frac{d\left[2d^2+9d-6- 2(d-1)\qq\right]}{3 (d-1)^2}$\\
    $b_7=\frac{d^3-2d^2-18d+12+2(d^2+4d-2)\qq}{d(d-1)(d+2)}$&$\theta_6=0$\\
    $b_8=\frac{2}{d-1}$&$\theta_7= 0$\\
\end{tabular}
\end{ruledtabular}
\end{table}

\subsection{Inelastic case}
The Burnett transport coefficients $a_i$ associated with the pressure tensor are plotted in Fig.\ \ref{fig1} as functions of $\alpha$ for two- and three-dimensional systems. In both cases, we have chosen  $\qq=\frac{1}{2}$,  which,  as said before, mimics the hard-sphere model. Similarly, Fig.\ \ref{fig2} shows the $\alpha$ dependence of the Burnett transport coefficients $b_i$ associated with the heat flux.

\begin{figure}
  \includegraphics[width=.45\textwidth]{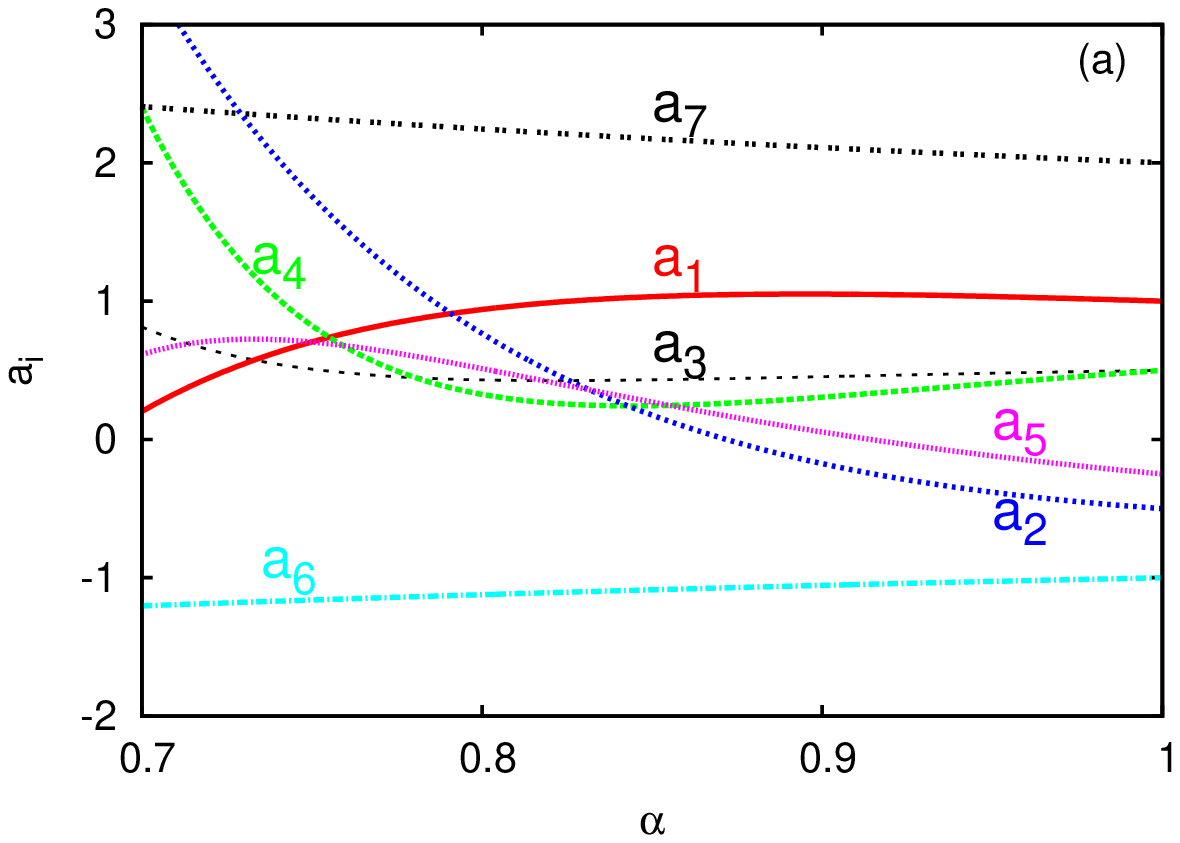}
  \includegraphics[width=.45\textwidth]{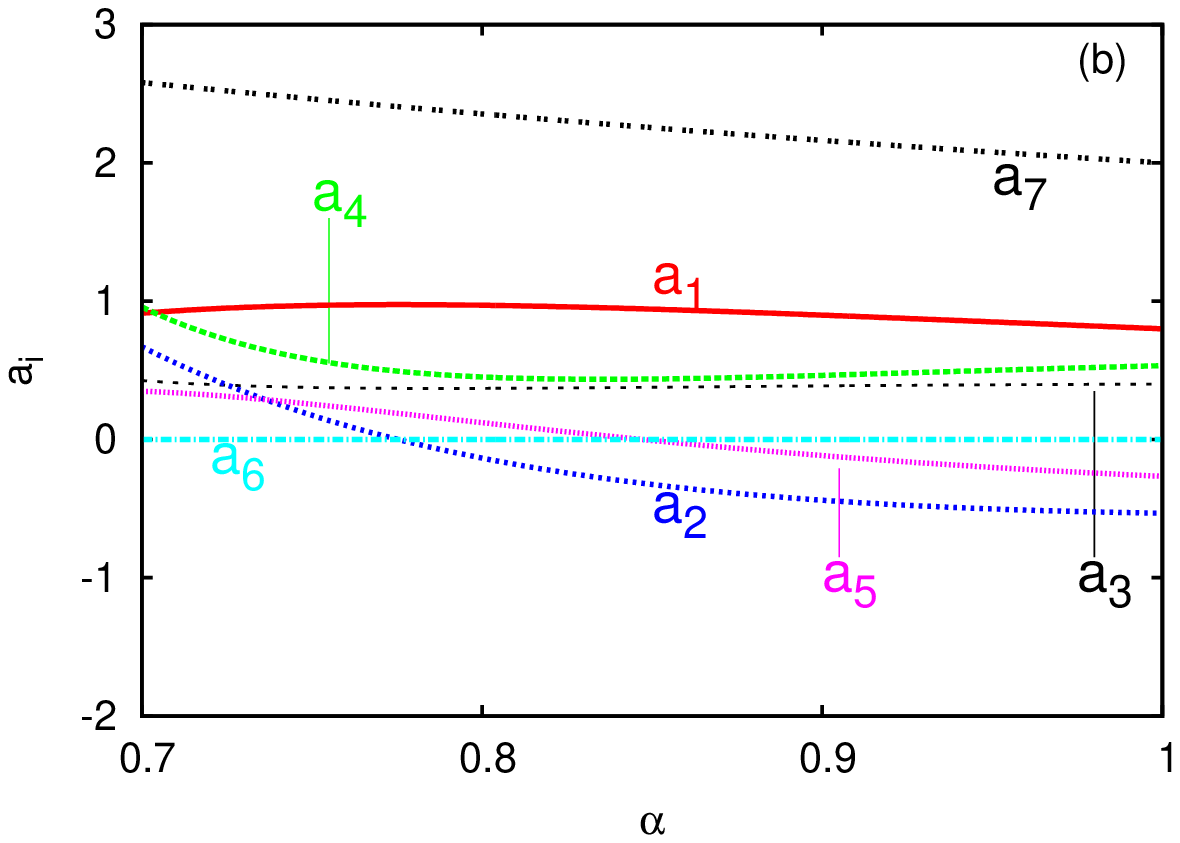}
\caption{IMM Burnett transport coefficients $a_1,\dots,a_7$ as functions of the coefficient of normal restitution  in the case $\qq=\frac{1}{2}$ for (a) $d=2$  and (b) $d=3$. }
\label{fig1}
\end{figure}

\begin{figure}
  \includegraphics[width=.45\textwidth]{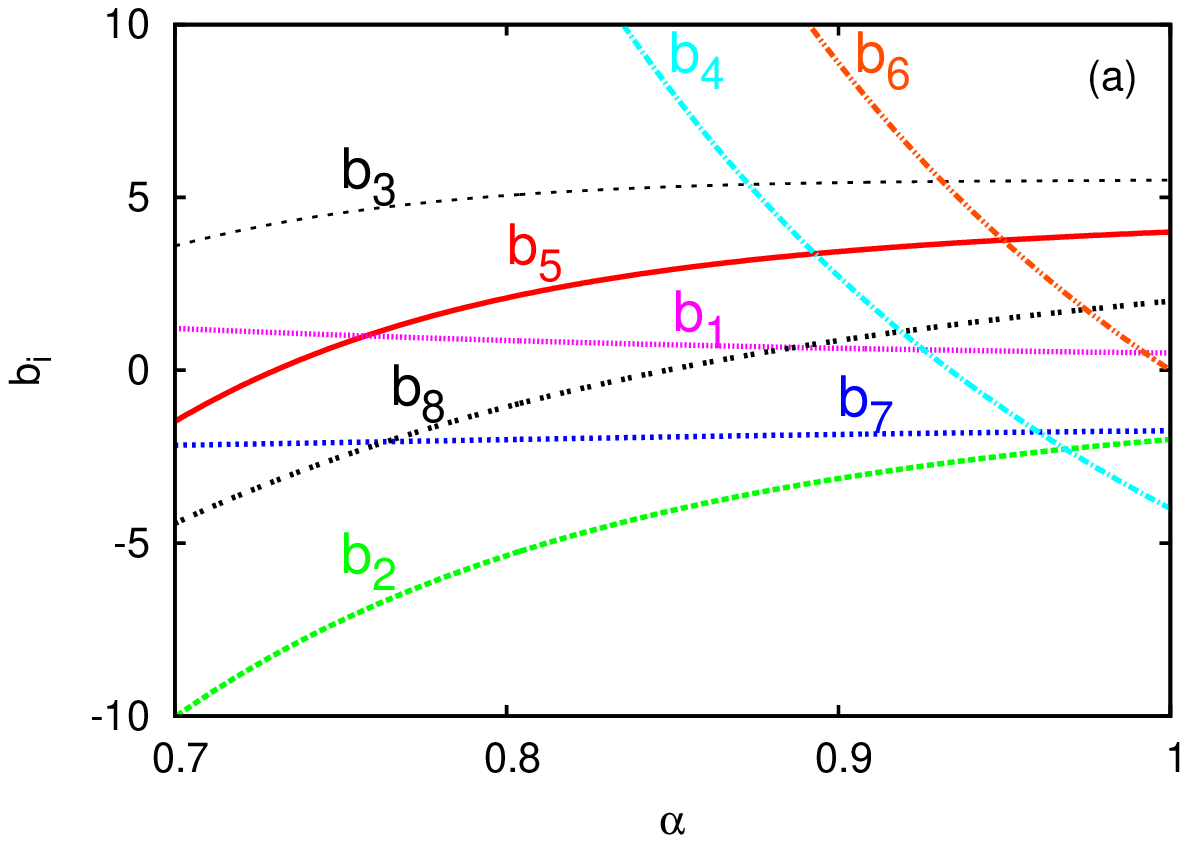}
  \includegraphics[width=.45\textwidth]{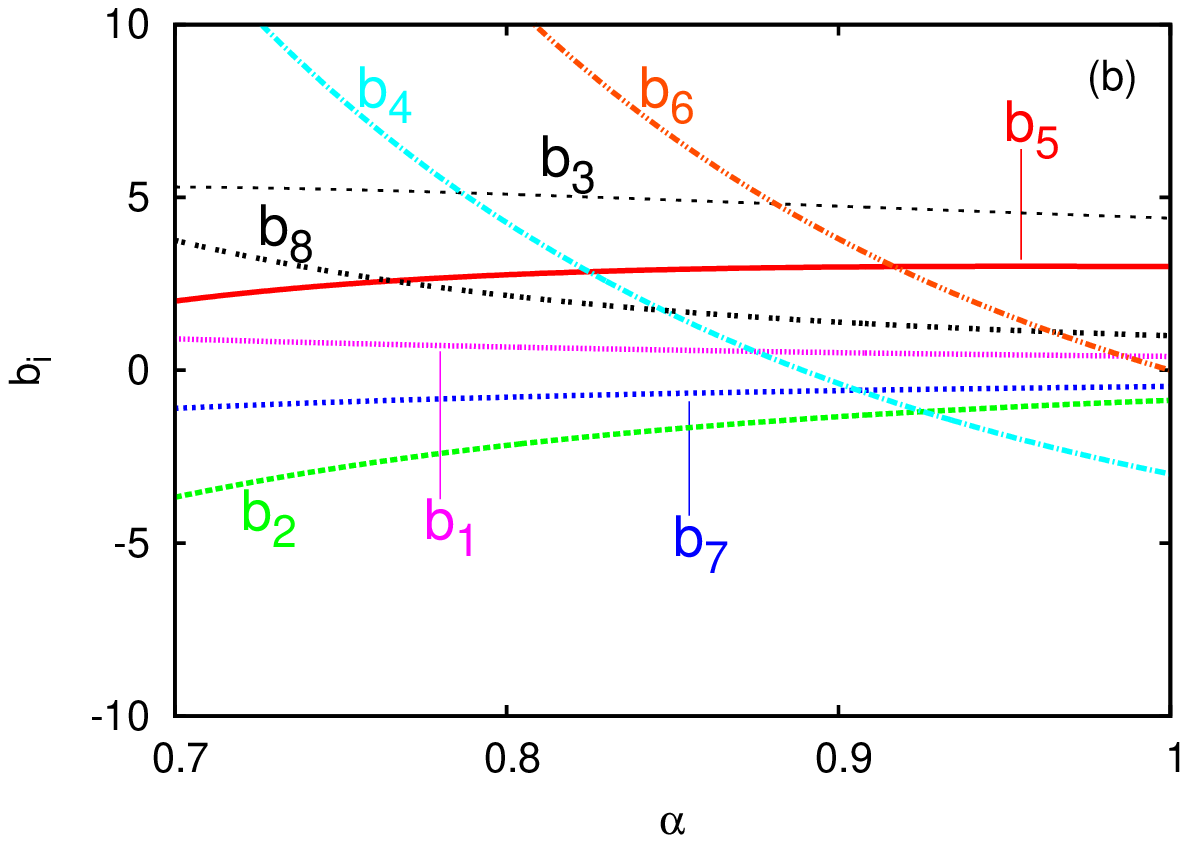}
\caption{IMM Burnett transport coefficients $b_1,\dots,b_8$ as functions of the coefficient of normal restitution  in the case $\qq=\frac{1}{2}$ for (a) $d=2$  and (b) $d=3$.}
\label{fig2}
\end{figure}

Within the range $0.7\leq \alpha\leq 1$, we observe that, whereas some coefficients ($a_3$, $a_6$, $b_1$, $b_3$, and $b_7$) exhibit a weak dependence on dissipation, other coefficients ($a_2$, $a_4$, $b_2$, $b_4$, and $b_6$) are quite sensitive to $\alpha$. The remaining coefficients ($a_1$, $a_5$, $a_7$, $b_5$, and $b_8$) present an intermediate behavior. It is especially interesting to note that the coefficient $b_6$, which vanishes in the elastic limit, grows very rapidly with increasing dissipation. In general, the impact of dissipation on the Burnett coefficients is  more significant for $d=2$ than for $d=3$.

It is known that the heat flux NS coefficients $\kappa^*$ and $\mu^*$ for the two- and three- dimensional IMM diverge, thus implying a breakdown of hydrodynamics, for very low values of $\alpha$ \cite{S03,BGM10,GS11}. As seen from Eq.\ \eqref{eq:28b}, the threshold for the divergence  occurs when $\nu_{2|1}^*=2\zeta^*$, i.e., at $\alpha=(4-d)/3d$, what corresponds to $\alpha=\frac{1}{3}$ and $\frac{1}{9}$ for $d=2$ and $3$, respectively.  While the Burnett coefficients $a_6$ and $a_7$ are always convergent, the remaining coefficients  may diverge. This is discussed in Appendix \ref{appC}. The regions in the $(\alpha,\qq)$ plane where the coefficients diverge are presented in Fig.\ \ref{fig3}.

\begin{figure}
  \includegraphics[width=.45\textwidth]{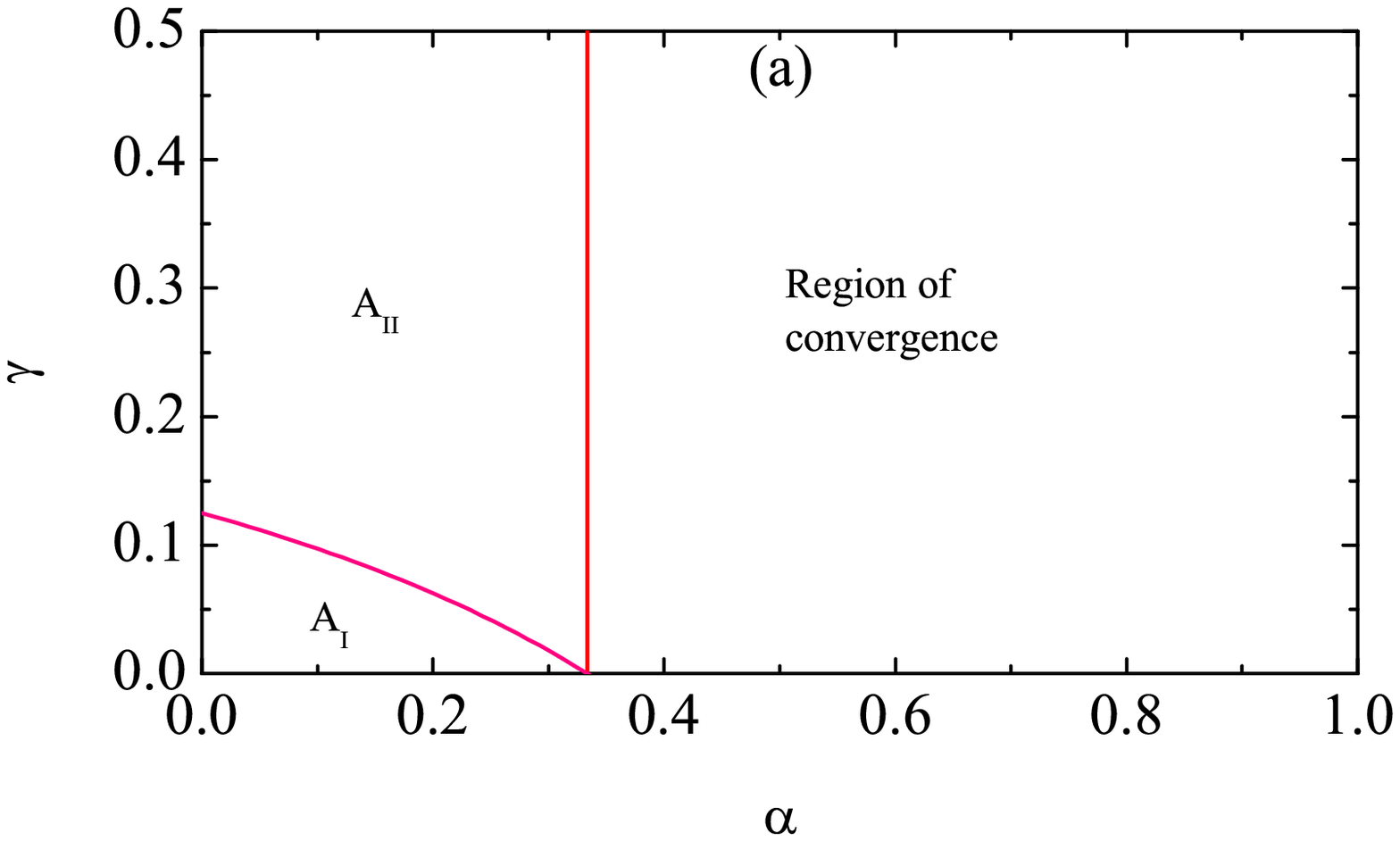}
  \includegraphics[width=.45\textwidth]{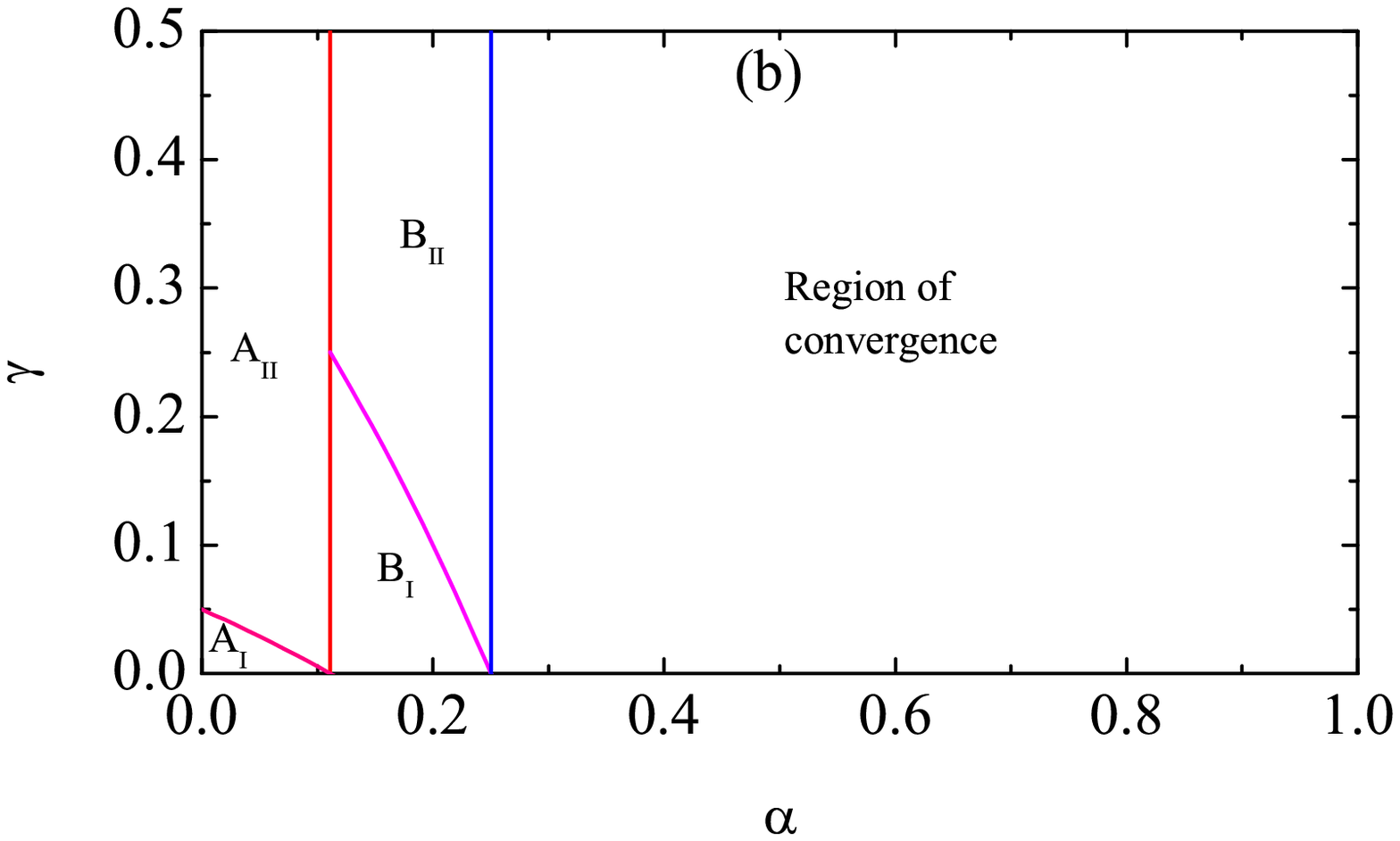}
  \caption{(a) Regions in the $(\alpha,\qq)$ plane where the IMM Burnett coefficients for a two-dimensional system diverge. The coefficient $b_1$ diverges in region $\text{A}_\text{I}$, while the coefficients $a_1$--$a_5$ and $b_2$--$b_8$ diverge in regions $\text{A}_\text{I}$ and $\text{A}_\text{II}$. (b)  Regions in the $(\alpha,\qq)$ plane where the Burnett coefficients for a three-dimensional system diverge. The coefficient $b_1$ diverges in region $\text{A}_\text{I}$, while the coefficients $b_2$--$b_8$ diverge in regions $\text{A}_\text{I}$ and $\text{A}_\text{II}$, the coefficients $a_1$ and $a_2$ diverge in regions $\text{A}_\text{I}$, $\text{A}_\text{II}$, and $\text{B}_\text{I}$, and the coefficients $a_3$--$a_5$ diverge in regions $\text{A}_\text{I}$, $\text{A}_\text{II}$,  $\text{B}_\text{I}$, and $\text{B}_\text{II}$.}
\label{fig3}
\end{figure}

{}In any case, from a practical point of view, all the NS and Burnett coefficients are well defined in the region of physical interest $\alpha\geq 0.5$.

\begin{figure}
  \includegraphics[width=.45\textwidth]{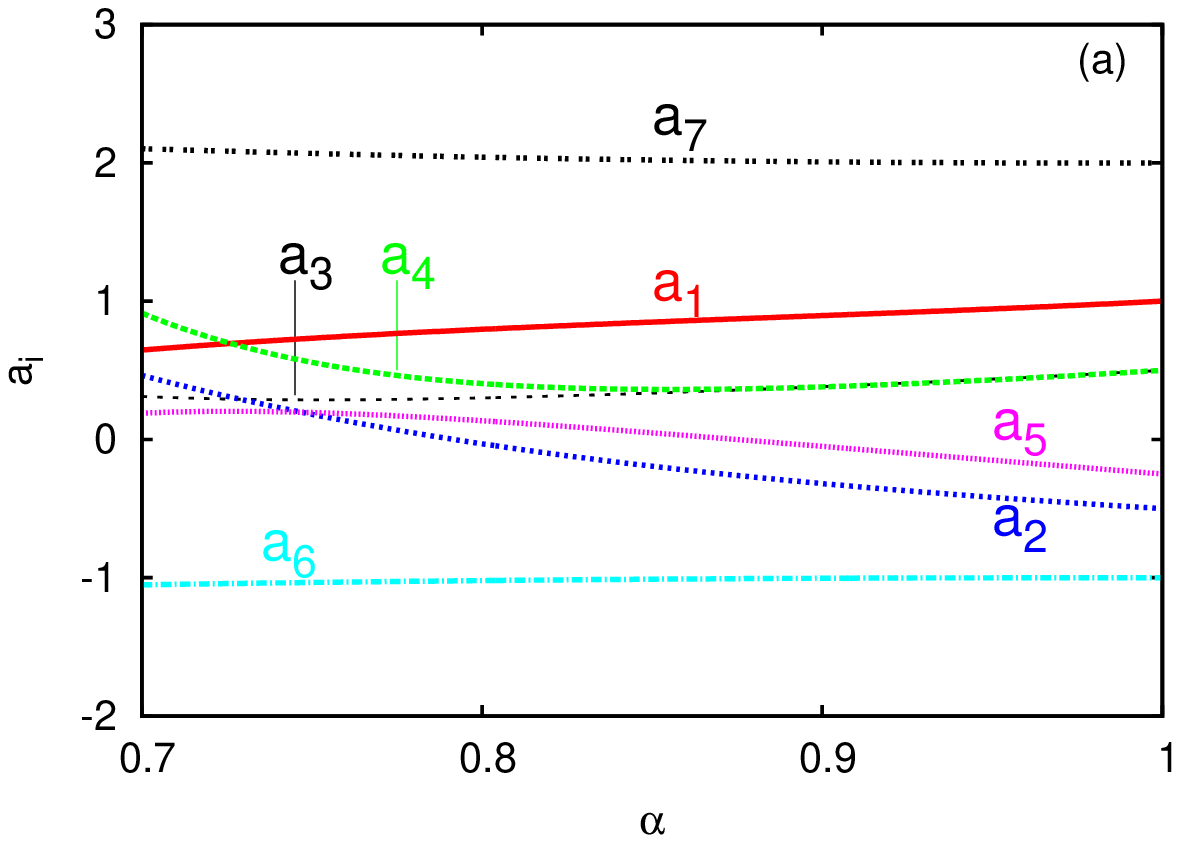}
  \includegraphics[width=.45\textwidth]{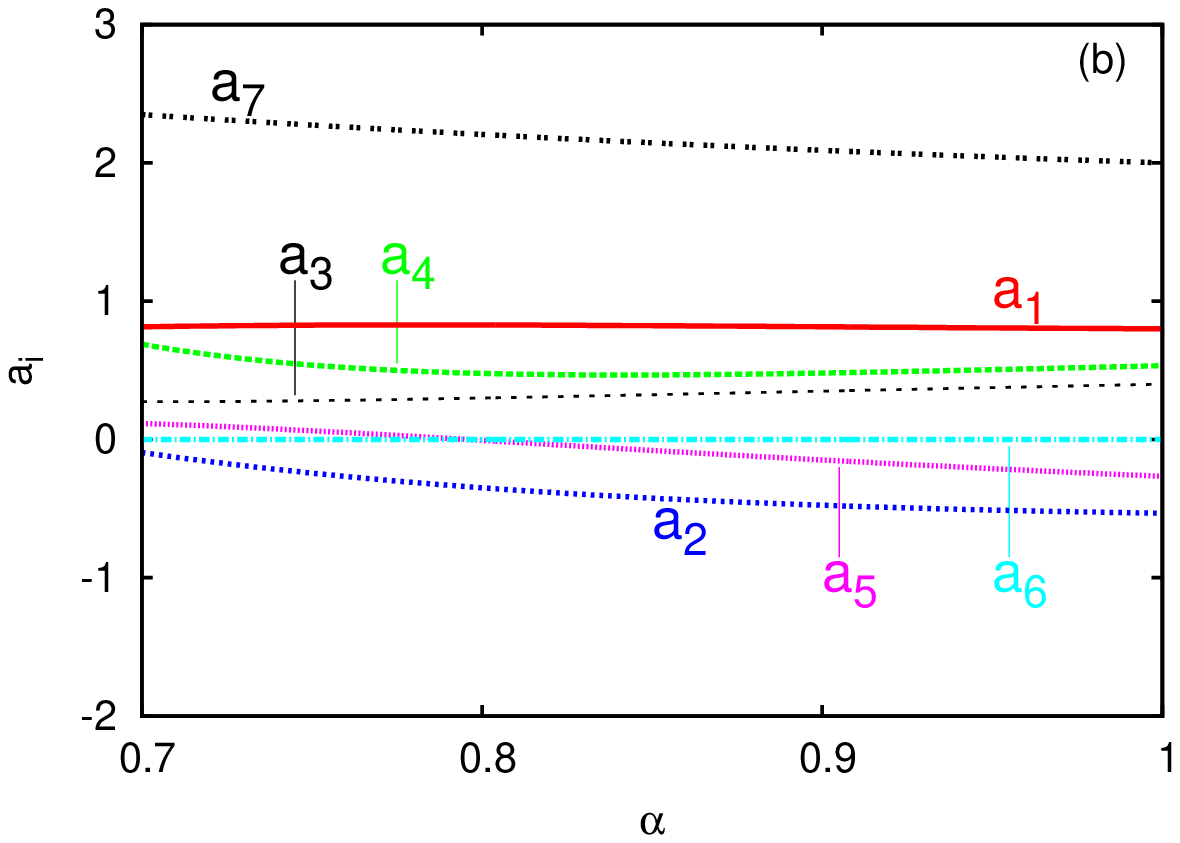}
\caption{Estimates of the IHSM Burnett transport coefficients $a_1,\dots,a_7$ as functions of the coefficient of normal restitution for (a) $d=2$  and (b) $d=3$. }
\label{fig4}
\end{figure}

\begin{figure}
  \includegraphics[width=.45\textwidth]{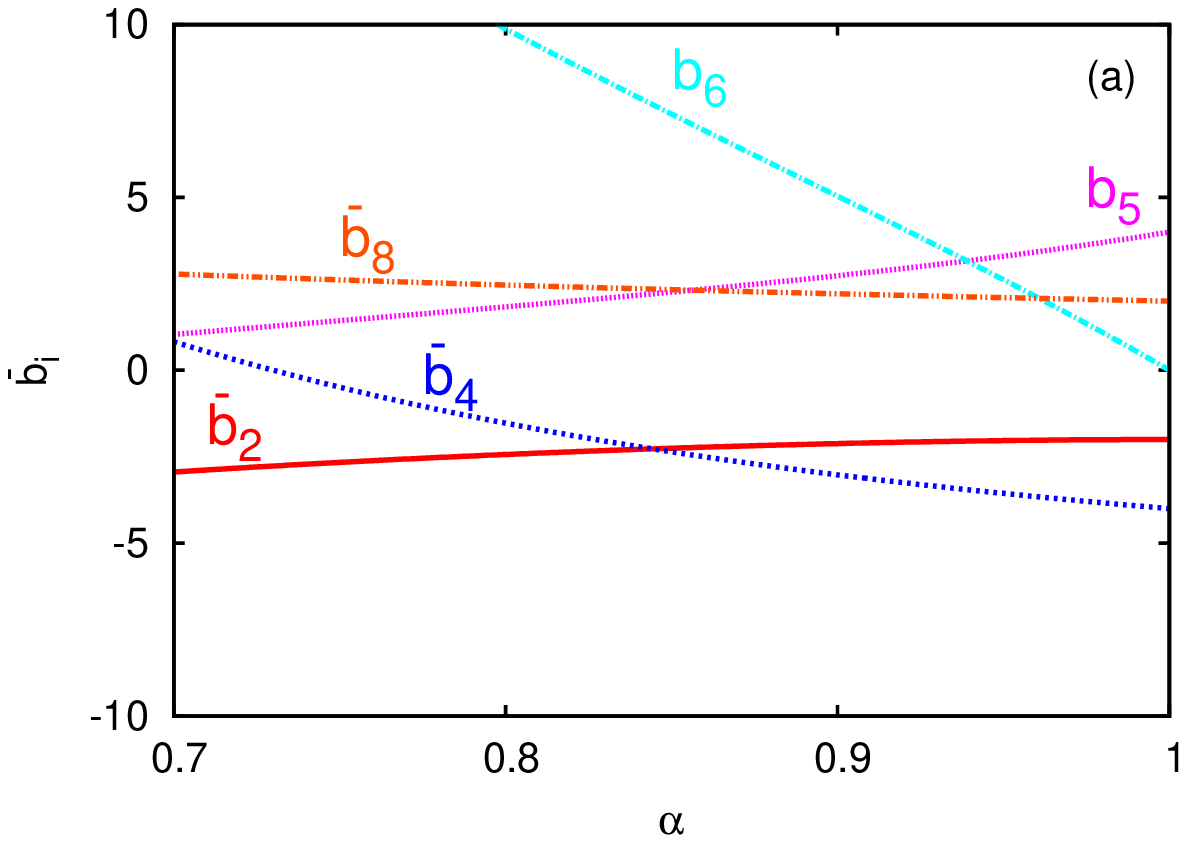}
  \includegraphics[width=.45\textwidth]{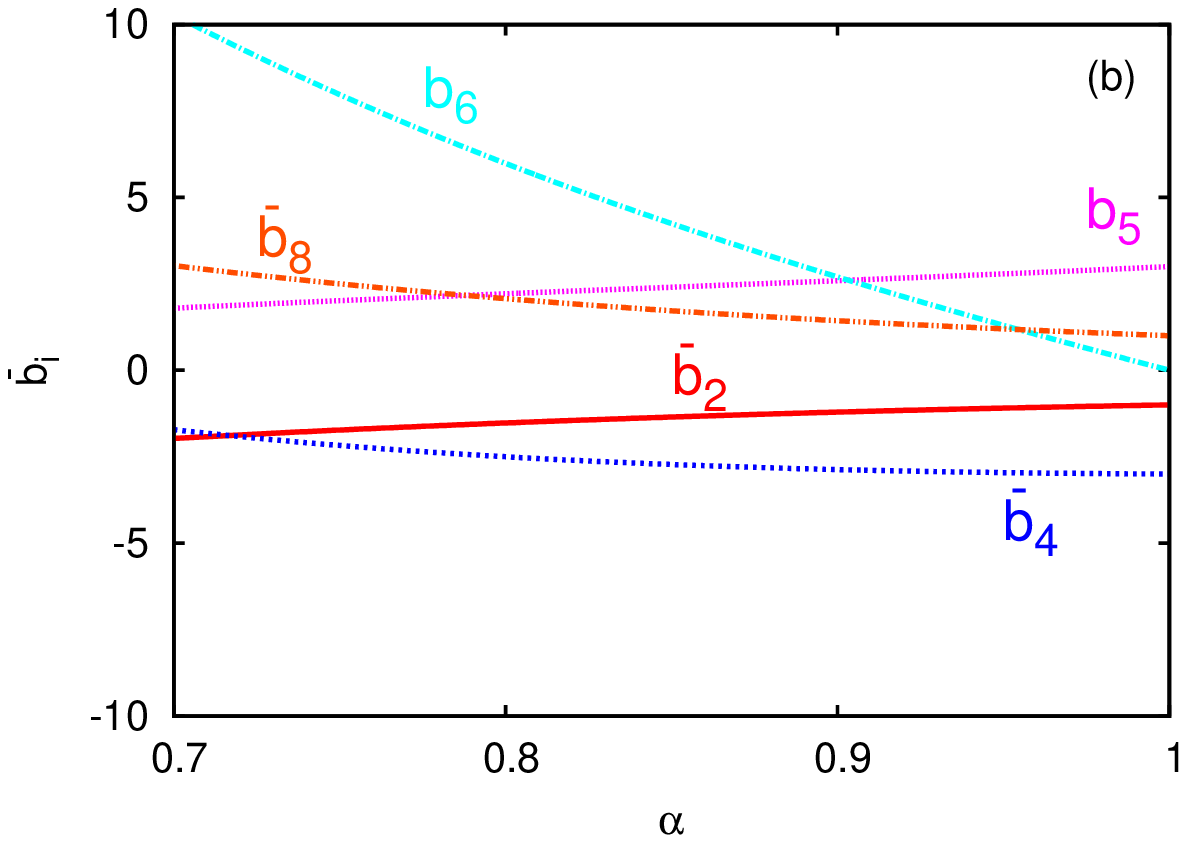}
\caption{Estimates of the IHSM Burnett transport coefficients $\overline{b}_2$, $\overline{b}_4$, ${b}_5$, ${b}_6$, and $\overline{b}_8$ as functions of the coefficient of normal restitution for (a) $d=2$  and (b) $d=3$. }
\label{fig5}
\end{figure}


\subsection{Estimates of IHSM coefficients}
\label{sec5C}
Although this paper is focused on the IMM, it is tempting to use the results derived here to \emph{estimate} the Burnett transport coefficients for the more realistic case of the IHSM. It is reasonable to expect that the mathematical structures of the constitutive equations \eqref{e33} and \eqref{e56} are essentially preserved in the IHSM case.

As said before, the structure of the NS coefficients, Eqs.\ \eqref{eq:28}--\eqref{eq:28c}, is exactly the same for both inelastic models, the differences lying in the $\alpha$ dependence of the cumulant $c$, cooling rate $\zeta^*$, and collision frequencies $\nu_{0|2}^*$ and $\nu_{2|1}^*$. Although the latter quantities are not exactly known for the IHSM, good estimates have been obtained from improved Sonine approximations \cite{vNE98,MS00,BDKS98,GSM07}. Their expressions are given in Appendix \ref{appD}.

We recall that, according to Eqs.\ \eqref{e40}--\eqref{e44}, the IMM Burnett coefficients $a_1$--$a_7$ associated with the pressure tensor depend on $\alpha$ \emph{only} through the four quantities $c$,  $\zeta^*$, $\nu_{0|2}^*$, and $\nu_{2|1}^*$. This suggests that  educated guesses for the corresponding IHSM Burnett coefficients can be obtained by inserting the corresponding IHSM values for  $c$,  $\zeta^*$, $\nu_{0|2}^*$, and $\nu_{2|1}^*$ into Eqs.\ \eqref{e40}--\eqref{e44} with $\qq=\frac{1}{2}$. The results are displayed in Fig.\ \ref{fig4}.
Comparison with Fig.\ \ref{fig1} shows  qualitatively similar behaviors, except that the influence of inelasticity is milder in the IHSM than in the IMM. {This is essentially related to the different types of high-velocity tails of the HCS distribution. While the tail is algebraic in the case of the IMM \cite{BK02,EB02a}, it has a stretched exponential form  in the case of the IHSM \cite{vNE98}}. As a matter of fact, we have checked that the only diverging coefficients (at $\alpha=0.046$ and $\alpha=0.015$ for $d=2$ and $d=3$, respectively) are $a_3$--$a_5$. This divergence takes place when $\nu_{0|2}^*=2\zeta^*$, but it cannot be discarded that the divergence would disappear if more accurate expressions for $\nu_{0|2}^*$ and $\zeta^*$ were used in the region of extreme inelasticity.

In the case of the Burnett coefficients $b_1$--$b_8$ associated with the heat flux, Eqs.\ \eqref{e62}--\eqref{CC2}, $b_5$ and $b_6$ depend on $\alpha$ only through $c$,  $\zeta^*$, $\nu_{0|2}^*$, and $\nu_{2|1}^*$, but the remaining ones include an extra dependence through the quantity $\psi$, which is unknown in the IHSM. On the other hand, the combinations,
\beq
\overline{b}_2\equiv b_2-\frac{d-2}{d}b_1=-\frac{\frac{2}{d}\kappa^*+\mu^*}{\nu_{2|1}^*-2(1-\qq)\zeta^*},
\eeq
\bal
\overline{b}_4&\equiv b_4-\frac{d(d+2)}{2(d-1)}\frac{\zeta^*}{\nu_{2|1}^*-(3-2\qq)\zeta^*}b_3\nn
&=\frac{A_2}{\nu_{2|1}^*-(3-2\qq)\zeta^*},
\eal
\bal
\overline{b}_8&\equiv b_8-\frac{d(d+2)}{2(d-1)}\frac{\zeta^*}{\nu_{2|1}^*-(3-2\qq)\zeta^*}b_7\nn
&=\frac{C_2}{\nu_{2|1}^*-(3-2\qq)\zeta^*},
\eal
do not include $\psi$ and thus can be estimated for the IHSM (with $\qq=\frac{1}{2}$).
The results for $\overline{b}_2$, $\overline{b}_4$, $b_5$, $b_6$, and $\overline{b}_8$ are plotted in Fig.\ \ref{fig5}. A comparison with a similar plot for the IMM (not shown) again exhibits qualitative similarities with a weaker dependence on inelasticity in the case of the IHSM.

\section{Conclusions}
\label{sec5}

The main objective of this paper was to derive the constitutive equations for the pressure tensor and the heat flux  of a granular gas by means of the CE method  up to second (Burnett) order in the hydrodynamic gradients, with explicit expressions for the corresponding transport coefficients. Given the formidable difficulties of the task, and in order to obtain results free from uncontrolled approximations, we used the IMM, which allowed us to achieve exact results. The final expressions apply to any value of the coefficient of normal restitution $\alpha$, any dimensionality $d$, and any value of the ``interaction'' parameter $\qq$. As a bonus, in the elastic limit ($\alpha=1$), our results provide the explicit forms of the Burnett transport coefficients of a classical gas for any number of dimensions (see Table \ref{table2}), which, to the best of our knowledge, had not been derived before.

It is interesting  to remark that the structure of the inelastic Burnett constitutive equations is more general than that of the elastic counterpart. While the terms involving second-order gradients are the same in both situations, some Burnett coefficients that are degenerate in the elastic case [$\varpi_2=\varpi_2'=\varpi_2''$ in Eq.\ \eqref{CCP} and $\theta_2=\theta_2'$ in Eq.\ \eqref{CCq}] become different when $\alpha\neq 1$.
In general, the dependence of the Burnett coefficients on inelasticity is far from being trivial (see Figs.\ \ref{fig1} and \ref{fig2}): While some coefficients tend to increase or decrease with increasing inelasticity, other ones are hardly sensitive to $\alpha$.

We also exploited the formal structure of the results for the IMM to obtain reasonable estimates of the Burnett coefficients for the IHSM. We plan to derive expressions for those coefficients by starting from the genuine Boltzmann equation for the IHSM and using similar Sonine approximations, as made before in the case of the NS coefficients. It will be instructive to use those expressions to assess the degree of reliability of the ones estimated here.
{Previous studies \cite{SG98} considered a double expansion in the spatial gradients and in the degree of inelasticity up to second order, so that the Burnett transport coefficients coincided with their elastic forms.}

Finally, it is worthwhile noting the potential usefulness of the Burnett-order  hydrodynamic equations, as compared to the NS equations, to  describe physical problems  where gradients are not small. In fact, this is the typical situation in granular fluids due to the coupling between inelasticity and gradients \cite{G03}. {On the other hand, some care must be taken since the Burnett equations, as noted in Sec.\ \ref{sec1}, need some kind of regularization to avoid artificial instabilities \cite{B82,UVG00,JS01,B04,B06,CKK07,G08a}.}

\acknowledgments
This work has been supported by the Spanish Government through Grant No.\ FIS2010-12587  and by the Junta de Extremadura (Spain) through Grant No.\ GRU10158, both partially financed by FEDER funds.

\appendix

\section{IHSM expressions for $c$, $\zeta^*$, $\nu_{2|0}^*$, and $\nu_{2|1}^*$}
\label{appD}
In the case of the IHSM, accurate estimates  are \cite{MS00,GSM07}
\beq
c= \frac{16(1-\alpha)(1-2\alpha^2)}{25+24d-\alpha (57-
8d)-2(1-\alpha)\alpha^2},
\label{D2.12}
\eeq
\beq
\zeta^*=\frac{d+2}{4d}(1-\alpha^2)\left(1+\frac{3c}{16}\right),
\label{x2.29}
\eeq
\begin{equation}
\label{D2.35}
\nu_{0|2}^*=\frac{(1+\alpha)\left[d+\frac{3}{2}(1-\alpha)\right]}{2d}
\left(1+\frac{7c}{16}\right),
\end{equation}
\bal
\label{D2.36}
\nu_{2|1}^*=&
\frac{1+\alpha}{8d}\left[\frac{11}{2}d+8-\frac{3}{2}\alpha(d+8)\right.\nn
&\left.+\frac{296+217d-3(160+11d)\alpha}{32}c\right].
\eal
These expressions are employed in Sec.\ \ref{sec5C} to estimate the Burnett coefficients in the IHSM.

\section{Evaluation of $P_{ij}^{(2)}$}
\label{appA}
Multiplying both sides of Eq.\ \eqref{eq:30} by $mV_iV_j$ and integrating over $\mn{v}$ one gets
\bal
\label{e24}
 \left(\partial_t^{(0)}+\nu_{0|2}\right)P_{ij}^{(2)}=& -m\int\, d\mn{
v}\; V_iV_j\left(\partial_t^{(1)}+\mn{ v}\cdot \boldsymbol{\nabla}
\right)f^{(1)} \nonumber \\& +\delta_{ij}\frac{2}{d}\left(\boldsymbol{\nabla} \cdot \mn{
q}^{(1)}+P_{k\ell}^{(1)}\nabla_\ell u_k \right),
\eal
where use has been made of the relation \eqref{eq:31b} and the collisional moment \cite{GS07}
\beq
m\int d\mn{v}\, V_i V_j J^{(2)}[f,f]=-\nu_{0|2}P_{ij}^{(2)},
\eeq
where $\nu_{r|s}=\nu_{r|s}^*\nu_0$ and $\nu_{0|2}^*$ is given by Eq.\ \eqref{eq:a1_6}.
The first term on the right-hand side of  Eq.\ \eqref{e24} can be easily evaluated with the result
\bal
\label{e26}
m\int &d\mn{ v}\; V_iV_j\left(\partial_t^{(1)}+\mn{
v}\cdot \boldsymbol{\nabla} \right)f^{(1)}=D_t^{(1)}P_{ij}^{(1)}+\nabla_k
Q_{ijk}^{(1)}  \nonumber \\
& +P_{ij}^{(1)}\boldsymbol{\nabla} \cdot \mn{
u}+P_{kj}^{(1)}\nabla_k u_i+P_{ki}^{(1)}\nabla_k u_j,
\eal
where $D_t^{(1)}=\partial_t^{(1)}+\mn{u}\cdot \boldsymbol{\nabla}$ is the material derivative  and the tensor $Q_{ijk}^{(1)}$ is defined as
\begin{equation}
  \label{e27}
  Q_{ijk}^{(1)}=m\int\; d\mn{ v}\; V_iV_jV_k f^{(1)}.
\end{equation}
We now evaluate separately $\nabla_k Q_{ijk}^{(1)}$ and $D_t^{(1)}P_{ij}^{(1)}$.


The NS quantity $Q_{ijk}^{(1)}$ can be  evaluated in a way similar to the evaluation of $\mn{q}^{(1)}$.
First, we multiply  Eq.\ \eqref{eq:23}  by $V_iV_jV_k$ and integrate over velocity. The result is
\bal
\label{eq:a2_2}
  & \partial_t^{(0)}Q_{ijk}^{(1)}+\nabla_\ell M_{ijk\ell}^{(0)}-\frac{p}{\rho}\left(\delta_{ij}\nabla_k p+\delta_{jk}\nabla_i p+\delta_{ik}\nabla_j p\right) \nonumber  \\ & \quad =m\int\; d\mn{ v}\; V_iV_jV_k J^{(1)}[f,f],
\eal
where
\bal
\label{eq:a2_3}
M_{ijk\ell}^{(0)}&=m\int\; d\mn{ v}\; V_iV_jV_k V_\ell
f^{(0)}\nn
&=\frac{pT}{m}(1+\cum)
\left(\delta_{ij}\delta_{k\ell}+\delta_{ik}\delta_{j\ell}+\delta_{i\ell}\delta_{jk}\right).
\eal
 The right-hand side of Eq.\ \eqref{eq:a2_2} can be explicitly evaluated as \cite{GS07}
\bal
\label{eq:a2_6}
& m\int\; d\mn{ v}\; V_iV_jV_k
J^{(1)}[f,f]=-\frac{3}{2}\nu_{0|2}Q_{ijk}^{(1)}\nn
& \quad + \frac{2}{d+2}\left(\frac{3}{2}\nu_{0|2}-\nu_{2|1}\right)\left(\delta_{ij}q_k^{(1)}
+\delta_{jk}q_i^{(1)}+\delta_{ik}q_j^{(1)}\right),
\eal
where $\nu_{2|1}^*$ is given by Eq.\ \eqref{eq:a1_6c}.
Substitution of Eq.\ \eqref{eq:a2_6}  into \eqref{eq:a2_2} yields
\bal
\label{eq:a2_8}
  \left(\partial_t^{(0)}+\frac{3}{2}\nu_{0|2}\right)  Q_{ijk}^{(1)}=&  -\nabla_\ell M_{ijk\ell}^{(0)} + \frac{p}{\rho}\left(\delta_{ij}\nabla_k p+\delta_{jk}\nabla_i p\right. \nn
  &\left.+\delta_{ik}\nabla_j p\right) + \frac{2}{d+2}\left(\frac{3}{2}\nu_{0|2}-\nu_{2|1}\right) \nn
  & \times \left(\delta_{ij}q_k^{(1)}
+\delta_{jk}q_i^{(1)}+\delta_{ik}q_j^{(1)}\right).
\eal
The solution to Eq.\ \eqref{eq:a2_8} has the form
\bal
\label{eq:a2_9}
Q_{ijk}^{(1)} =&-a_Q
(\delta_{ij}\nabla_k \ln p+  \delta_{jk}\nabla_i \ln p+\delta_{ik}\nabla_j \ln p) \nonumber  \\
& - b_Q(\delta_{ij}\nabla_k \ln T  + \delta_{jk}\nabla_i \ln T+\delta_{ik}\nabla_j \ln T),
\eal
where the coefficients $a_Q$ and $b_Q$ are  determined by consistency. They can be easily
obtained by taking into account the identity $q_k^{(1)}=\frac{1}{2}Q_{iik}^{(1)}$ with the result
\beq
 a_Q=\frac{2}{d+2}n\mu,\quad
  b_Q=\frac{2}{d+2}(T\kappa-n\mu).
\end{equation}
Thus, the gradient of $Q_{ijk}^{(1)}$ is
\begin{widetext}
\bal
\label{eq:a2_12}
\nabla_k
Q_{ijk}^{(1)}=&-\frac{2}{d+2}\frac{n\mu}{p}\left(\delta_{ij}\nabla^2p+
2\nabla_i\nabla_jp\right)  -\frac{2}{d+2}\left(\kappa-\frac{n\mu}{T}\right)
\left(\delta_{ij}\nabla^2T+ 2\nabla_i\nabla_jT\right) \nn
&-\frac{2}{d+2}\frac{n\mu}{p{T}}\Big\{(2-\qq)\delta_{ij}\left(\boldsymbol{\nabla} p\right)
\cdot \left(\boldsymbol{\nabla} T \right)  +(2-\qq)\left[\left(\nabla_ip\right)\left(\nabla_jT\right)+\left(\nabla_jp\right)\left(\nabla_iT\right)\right]   -\frac{T}{p}\delta_{ij}(\boldsymbol{\nabla} p)^2  \nn
&-2\frac{T}{p}\left(\nabla_ip\right)\left(\nabla_jp\right)\Big\}-\frac{2}{d+2}T^{-1}\left(\kappa-\frac{n\mu}{T}\right)(1-\qq)  \left[\delta_{ij}(\boldsymbol{\nabla} T)^2+ 2\left(\nabla_iT\right)\left(\nabla_jT\right)\right].
\eal

Now we turn to the evaluation of $D_t^{(1)} P_{ij}^{(1)}$. Using Eq.\ \eqref{eq:26} of the pressure tensor at NS order and the balance equations \eqref{eq:24}--\eqref{eq:24b}, one finds
\bal
\label{eq:a2_13}
 D_t^{(1)} P_{ij}^{(1)}=&\frac{2}{d}(1-\qq)\eta \left(\boldsymbol{\nabla} \cdot \mn{
u}\right)\left(\nabla_iu_j+\nabla_ju_i-\frac{2}{d}\delta_{ij}\boldsymbol{\nabla} \cdot
\mn{ u}\right) +\eta \left\{\nabla_i\left(\frac{1}{\rho}\nabla_j
p\right)+\nabla_j\left(\frac{1}{\rho}\nabla_i
p\right)  +\left(\nabla_iu_k\right)\left(\nabla_ku_j\right)\right.  \nn
  & \left. +\left(\nabla_j u_k\right)\left(\nabla_ku_i\right)-\frac{2}{d}\delta_{ij}\left[\boldsymbol{\nabla} \cdot
\left(\frac{1}{\rho}\boldsymbol{\nabla} p\right)+\left(\nabla_\ell u_k\right)\left(\nabla_k
u_\ell\right)\right]\right\}.
\eal

Substitution of Eqs.\ \eqref{eq:a2_12} and \eqref{eq:a2_13} into  Eq.\ \eqref{e24} yields
 \bal
\label{e31}
\Bigl(\partial_t^{(0)} +\nu_{0|2} \Bigr)P_{ij}^{(2)}=&
c_{P,1}\left(\nabla_i\nabla_jT-\frac{1}{d}\delta_{ij}\nabla^2T\right)
+ c_{P,2}\left(\nabla_i\nabla_j p- \frac{1}{d}\delta_{ij}\nabla^2p\right)
+c_{P,3}\left[(\nabla_iT)(\nabla_jT)-\frac{1}{d}\delta_{ij}(\boldsymbol{\nabla} T)^2\right]\nn
&  + c_{P,4} \left[(\nabla_ip)( \nabla_jp)- \frac{1}{d}\delta_{ij}{(\boldsymbol{\nabla} p)^2}\right]   +c_{P,5} \left[(\nabla_i T)(\nabla_j p)+(\nabla_i p)(\nabla_jT)-\frac{2}{d}\delta_{ij}(\boldsymbol{\nabla} p )\cdot (\boldsymbol{\nabla} T)\right]  \nonumber \\
&+c_{P,6} D\left(D_{ij}-\frac{1}{d}\delta_{ij} D\right) +c_{P,7} {\left[D_{ik}D_{kj}-\omega_{ik}\omega_{kj}-\frac{1}{d}\delta_{ij} \left(D_{lk}D_{kl}-\omega_{lk}\omega_{kl}\right)-D_{ik}\omega_{kj}-D_{jk}\omega_{ki}\right]},
\eal
\end{widetext}
where the coefficients $c_{P,i}$ are
 \beq
  c_{P,1}=\frac{4}{d+2} \left(\kappa-\frac{n\mu}{T}\right),
   \eeq
   \beq
 c_{P,2}=- p{c_{P,4}}=\frac{4}{d+2}\frac{n\mu}{p}-\frac{2\eta}{\rho},
 \eeq
\begin{equation}
\label{eq:a2_20a}
 c_{P,3}=-\frac{4}{d+2}(\qq-1)T^{-1} \left(\kappa-\frac{n\mu}{T}\right),
  \eeq
 \beq
 c_{P,5}=\frac{2}{d+2}(2-\qq)\frac{n\mu}{Tp}-\eta \rho^{-1}T^{-1},
  \eeq
 \beq
  c_{P,6}=-\frac{2}{d}\eta (4-d-2\qq),
   \eeq
 \beq
 c_{P,7}= 2\eta.
\eeq

The structure of Eq.\ \eqref{e31} shows that the constitutive equation for $P_{ij}^{(2)}$ has the form \eqref{e33}, where the dimensionless coefficients $a_i$ can be determined by inserting Eq.\ \eqref{e33} into Eq.\ \eqref{e31} and equating coefficients of the same type of gradients. After tedious algebra one finally gets Eqs.\ \eqref{e40}--\eqref{e42}.

It is interesting to remark that, while $ c_{P,2}=- p{c_{P,4}}$, one has $a_2\neq -a_4$ (except in the elastic limit). This is due to the different action of the operator $\partial_t^{(0)}$ on $p^{-1}(\nabla_i p)(\nabla_j p)$ and $\nabla_i\nabla_j p$.

\section{Evaluation of $\mn{q}^{(2)}$}
\label{appB}
The evaluation of $\mn{q}^{(2)}$ proceeds along similar lines as in the case of $P_{ij}^{(2)}$. First, by multiplying both sides of Eq.\ \eqref{eq:30} by $\frac{m}{2}V^2V_i$ and integrating over velocity, one obtains\bal
\label{e46}
\left(\partial_t^{(0)}+\nu_{2|1}\right)q_{i}^{(2)}=&-\frac{m}{2}\int\; d\mn{
v}\; V^2V_i\left(\partial_t^{(1)}+\mn{ v}\cdot \boldsymbol{\nabla}
\right)f^{(1)} \nonumber \\ & +\frac{d+2}{2}\frac{p}{\rho}\nabla_j P_{ij}^{(1)},
\eal
where use has been made of the relation \eqref{eq:31a} and \cite{GS07}
\beq
\frac{m}{2}\int d\mn{v}\, V^2 V_i J^{(2)}[f,f]=-\nu_{2|1}q_i^{(2)}.
\eeq

The first term on the right-hand side of Eq.\ \eqref{e46} becomes
\begin{widetext}
  \beq
\label{e48}
  \frac{m}{2}\int d\mn{ v} V^2V_i \left(\partial_t^{(1)}+\mn{
v}\cdot \boldsymbol{\nabla} \right)f^{(1)}=D_t^{(1)}q_{i}^{(1)}+\nabla_j R_{ij}^{(1)}
+ Q_{ijk}^{(1)}\nabla_k u_j
+q_{i}^{(1)}\boldsymbol{\nabla} \cdot \mn{
u}+q_{j}^{(1)}\nabla_j u_i-\rho^{-1}P_{ij}^{(1)}\nabla_jp,
\eeq
\end{widetext}
where
\begin{equation}
  \label{e48a}
  R_{ij}^{(1)}=\frac{m}{2}\int\; d\mn{ v}\,V^2 V_iV_j
f^{(1)}.
\end{equation}

In order to evaluate $R_{ij}^{(1)}$, let us multiply  both sides of Eq.\ \eqref{eq:23} by $\frac{m}{2}V^2 V_iV_j$ and integrate over velocity to obtain
\bal
\label{eq:a2_27}
 \partial_t^{(0)}R_{ij}^{(1)}=& -\frac{pT}{m} ({d+4})(1+\cum) \left(D_{ij}-\frac{D}{d}\delta_{ij}\right) \nonumber \\
&+\frac{m}{2}\int d\mn{v}\, V^2 V_i V_j J^{(1)}[f,f].
\eal
The collision integral is \cite{GS07}
\bal
\label{eq:a2_22}
\frac{m}{2}\int d\mn{ v}\,  V^2V_iV_j
 J^{(1)}[f,f] =&-\nu_{2|2}R_{ij}^{(1)}{+
\frac{dp}{2\rho}\lambda P_{ij}^{(1)}}\nn
 &+{\frac{m}{2}}
\delta_{ij}(\nu_{2|2}-\nu_{4|0})M_{4|0}^{(1)},
\eal
where $\nu_{2|2}^*$ and $\lambda^*=\lambda/\nu_0$ are given by Eqs.\ \eqref{eq:62c} and \eqref{eq:a1_7}, respectively, $\nu_{4|0}$ can be found in Ref.\ \cite{GS07} but will not be needed here, and
\beq
M_{4|0}^{(1)}=\int d\mn{v}\, V^4 f^{(1)}(\mn{v}).
\eeq
However, as said in Sec.\ \ref{sec3b}, $M_{4|0}^{(1)}=0$ [see Eq.\ \eqref{eq:25}].
Consequently, Eq.\ \eqref{eq:a2_27} becomes
\bal
\label{eq:a2_27b}
 \Big(\partial_t^{(0)}+ \nu_{2|2}\Big)R_{ij}^{(1)}=& {-\frac{pT}{m}\left[(d+4)(1+\cum)+\frac{d}{p}\lambda\eta\right]}
\nonumber \\ & {\times \left(D_{ij}-\frac{D}{d}\delta_{ij}\right).}
\eal
Its solution is
\begin{equation}
\label{eq:a2_29}
R_{ij}^{(1)}=-\frac{T\eta_0}{m}\psi
\left(D_{ij}-\frac{D}{d}\delta_{ij}\right),
\end{equation}
$\psi$ being given by Eq.\ \eqref{eq:62a}.
The divergence of the tensor $R_{ij}^{(1)}$ is
\bal
\label{eq:a2_31}
 \nabla_j  R_{ij}^{(1)}=& -\frac{\eta_0}{m}(2-\qq)\psi  \left( D_{ij}-\frac{1}{d}\delta_{ij}D\right)\nabla_jT \nonumber\\
& -\frac{T\eta_0}{2m}\psi
 \left(\nabla^2 u_i+\frac{d-2}{d}\nabla_iD\right).
\eal

{From} Eq.\ \eqref{eq:26a}  and the balance equations \eqref{eq:24}--\eqref{eq:24b}, one gets
\bal
\label{eq:a2_30}
D_t^{(1)} q_{i}^{(1)}  =&\frac{2}{d}(2-\qq)\left[\frac{n\mu}{p}D\nabla_ip
 +\left(\kappa-\frac{n\mu}{T}
\right)D\nabla_iT\right]  \nonumber \\
& +\left(\frac{2}{d}\kappa+\frac{n\mu}{T}\right) T\nabla_iD +\frac{n\mu}{p}(\nabla_iu_j)(\nabla_jp)
\nonumber \\
& +\left(\kappa-\frac{n\mu}{T}\right)(\nabla_iu_j)(\nabla_jT).
\eal

Using Eqs.\ \eqref{eq:a2_31} and \eqref{eq:a2_30},  Eq.\ \eqref{e46} reduces to
\bal
\label{e55}
\Big(\partial_t^{(0)}+\nu_{2|1}\Big)q_{i}^{(2)}=&
c_{q,1} \nabla^2u_i
+ c_{q,2} \nabla_iD+c_{q,3}D_{ij}\nabla_jT
\nn
&
+ c_{q,4} D_{ij}\nabla_jp
 + c_{q,5} \omega_{ij}\nabla_jT
 \nn
 &+c_{q,6} \omega_{ij}\nabla_j p
 +c_{q,7} D\nabla_iT   + c_{q,8} D\nabla_ip ,
\eal
where the coefficients $c_{q,i}$ are
\beq
c_{q,1}=\frac{T\eta_0}{2m}\left[\psi-(d+2)\eta^*\right],
\eeq
\beq
c_{q,2}= \frac{d-2}{d}\frac{T\eta_0}{2m}  \left[\psi -(d+2)\eta^* \right]-\frac{2}{d}T\kappa-{n\mu},
\eeq
\beq
 c_{q,3}= (d+2)\frac{\eta_0}{m}\Bigg[ \frac{2-\qq}{d+2}\psi  -(1-\qq)\eta^*\Bigg] +\frac{4}{d+2} \left(\kappa-\frac{n\mu}{T}\right),
\eeq
\beq
c_{q,4}= \frac{4}{d+2}\frac{n\mu}{p}-\frac{2\eta}{\rho},
\eeq
\beq
c_{q,5}=2\left(\kappa-\frac{n\mu}{T}\right), \quad
c_{q,6}={2\frac{n \mu}{p}},
\eeq
\bal
c_{q,7}=& -(d+2)\frac{\eta_0}{d m}\left[\frac{2-\qq}{d+2}\psi-(1-\qq)\eta^*\right] \nn
& +\frac{d^2-8+2\qq(d+2)}{d(d+2)}\left(\kappa-\frac{n\mu}{T}\right),
\eal
\begin{equation}
\label{eq:a2_35}
   c_{q,8}= \frac{2\eta}{d\rho}+ \frac{n\mu}{p}\frac{d^2-8+2\qq(d+2)}{d(d+2)}.
\end{equation}

Equation \eqref{e55} shows that the constitutive equation for $q_{i}^{(2)}$ has the structure \eqref{e56}. The dimensionless coefficients $b_i$ are obtained by inserting Eq.\ \eqref{e56} into Eq.\ \eqref{e55}. The final results are displayed by Eqs.\ \eqref{e62}, \eqref{e61}, and \eqref{b3b4}--\eqref{b7b8}.

{

{\section{Non-Newtonian uniform shear and uniform longitudinal flows\label{nonNewt}}}
{\subsection{Unsteady uniform shear flow}}
The uniform shear flow (USF) is an incompressible flow characterized by uniform density and temperature ($\boldsymbol{\nabla} n=\boldsymbol{\nabla} T=0$) and a uniform shear rate, i.e., $\nabla_iu_j=a^*\nu_0 \delta_{iy}\delta_{jx}$ with $\boldsymbol{\nabla} a^*=0$.

Proceeding in a way analogous to the case of the Boltzmann equation for the IHSM in the Grad approximation \cite{SGD04} or with a simple kinetic model \cite{AS07,S08,S08a,AS12}, it is possible to eliminate time in favor of the reduced shear rate $a^*$ to obtain a coupled set of two equations for the rheological dependence of $P_{xy}^*$ and $P_{yy}^*$ on $a^*$, where
$P_{ij}^*=P_{ij}/p$. In the case of our generalized IMM model, those two equations are exact and read \cite{G07}
\beq
\gamma\left(\frac{2a^*}{d}P_{xy}^*+\zeta^*\right)\frac{\partial{P}_{xy}^*}{\partial a^*}=-P_{yy}^*+\frac{2}{d}P_{xy}^{*2}-\frac{\nu_{0|2}^*-\zeta^*}{a^*}P_{xy}^*,
\label{3.12}
\eeq
\beq
\gamma\left(\frac{2a^*}{d}P_{xy}^*+\zeta^*\right)\frac{\partial{P}_{yy}^*}{\partial a^*}=\frac{2}{d}P_{yy}^*P_{xy}^*-\frac{\nu_{0|2}^*-\zeta^*}{a^*}\left(P_{yy}^*-1\right).
\label{3.13}
\eeq

The numerical solution of this set of equations provides $P_{xy}^*(a^*)$ and $P_{yy}^*(a^*)$  for any $a^*$ \cite{SG07}. Here, however, we are interested in the analytical results to second order in $a^*$. In that case,
\beq
P_{xy}^*(a^*)=-\eta^* a^*+\mathcal{O}(a^{*3}),
\label{Pxy}
\eeq
\beq
P_{yy}^*(a^*)=1+\frac{\Psi}{d}a^{*2}+\mathcal{O}(a^{*4}),
\label{Pyy}
\eeq
where $\eta^*$ is the (reduced) NS shear viscosity and $\Psi$ is a viscometric function. Inserting Eqs.\ \eqref{Pxy} and \eqref{Pyy} into Eqs.\ \eqref{3.12} and \eqref{3.13}, and equating terms of the same order in $a^*$ one recovers Eq.\ \eqref{eq:28} and obtains $\Psi=a_7$, where $a_7$ is given by Eq.\ \eqref{e41}. It is straightforward to check that Eq.\ \eqref{e33}, when particularized to the USF conditions, indeed yields Eq.\ \eqref{Pyy} with $\Psi=a_7$.

{\subsection{Unsteady uniform longitudinal flow}}
The uniform longitudinal flow (ULF) shares with the USF the conditions $\boldsymbol{\nabla} n=\boldsymbol{\nabla} T=0$, but it is a compressible  flow because now $\nabla_iu_j=a^*\nu_0 \delta_{ix}\delta_{jx}$, $\boldsymbol{\nabla} a^*=0$
 \cite{S08,S08a,S09,AS12}.
Eliminating again time in favor of the reduced strain (or longitudinal rate) $a^*$ it is possible to find a closed equation for $P_{xx}^*(a^*)$:
\bal
\gamma\left(\frac{2a^*}{d}P_{xx}^*+\zeta^*\right)\frac{\partial{P}_{xx}^*}{\partial a^*}=&-2P_{xx}^*\left(1-\frac{P_{xx}^{*}}{d}\right)\nn
&-\frac{\nu_{0|2}^*-\zeta^*}{a^*}\left(P_{xx}^*-1\right).
\label{3.12ULF}
\eal
To second order in $a^*$ one can write
\beq
P_{xx}^*=1-2\frac{d-1}{d}\eta^* a^*+\frac{d-1}{d}\Phi a^{*2}+\mathcal{O}(a^{*3}),
\label{Pxx}
\eeq
where $\eta^*$ is again the (reduced) NS shear viscosity but $\Phi$ is a viscometric function different from $\Psi$. Substitution of Eq.\ \eqref{Pxx} into Eq.\ \eqref{3.12ULF} allows one to recover Eq.\ \eqref{eq:28} and obtain $\Phi=a_6+a_7$, where $a_6$ and $a_7$ are given by Eqs.\ \eqref{e40} and \eqref{e41}, respectively. As before, it can be checked that Eq.\ \eqref{Pxx} is indeed equivalent to Eq.\ \eqref{e33} particularized to the ULF conditions.

}

\section{Divergence of the Burnett coefficients}
\label{appC}
In this appendix we analyze the regions in the $(\alpha,\qq)$ plane where the Burnett coefficients for the IMM diverge.

It can be checked that $\nu_{0|2}^*-\zeta^*>0$ for all $\alpha$, so that $\eta^*$, $a_6$, and $a_7$ are finite [cf.\ Eqs.\ \eqref{eq:28}, \eqref{e40}, and \eqref{e41}]. Next, from Eqs.\ \eqref{a2} and \eqref{a4} one finds that the divergence threshold $\alpha_\thr^{(a_1,a_2)}$ for $a_1$ and $a_2$ takes place either  at $\alpha=(4-d)/3d$  or when $\Delta=0$, i.e., $\nu_{0|2}^*-(2-\qq)\zeta^*=0$, whatever comes first. Therefore,
\beq
\alpha_\thr^{(a_1,a_2)}=\begin{cases}
  \frac{4-d}{3d},&\frac{d-2}{2(d-1)}\leq \qq\leq \frac{1}{2},\\
  \frac{1-(d+2)\qq/2}{d+1-(d+2)\qq/2},&0\leq \qq\leq \frac{d-2}{2(d-1)}.
\end{cases}
\label{C1}
\eeq
Regarding the coupled coefficients $a_3$, $a_4$, and $a_5$, the determinant of the matrix $\mathsf{L}$ is  $(\nu_{0|2}^*-2\zeta^*)\Delta$, so that the divergence of $\kappa^*$ and $\mu^*$ is preempted by $\nu_{0|2}^*-2\zeta^*=0$. Thus,
\beq
\alpha_\thr^{(a_3,a_4,a_5)}=\frac{1}{d+1}.
\label{C2}
\eeq

Now we turn to the heat flux Burnett coefficients.
It can be checked that $\nu_{2|2}^*-2\zeta^*>0$ for all $\alpha$, so that $\psi$ is finite.
As a consequence, the threshold value $\alpha_\thr^{(b_1)}$ for $b_1$ occurs when $\nu_{2|1}^*-2(1-\qq)\zeta^*=0$ [cf.\ Eq.\ \eqref{e62}]. This implies
\beq
\alpha_\thr^{(b_1)}=\frac{4-d-4(d+2)\qq}{3d-4(d+2)\qq}.
\label{C3}
\eeq
In the case of $b_2$ [cf.\ Eq.\ \eqref{e61}], its divergence is due to that of $\kappa^*$ and $\mu^*$, i.e.,
\beq
\alpha_\thr^{(b_2)}=\frac{4-d}{3d}.
\label{C4}
\eeq
Finally, the determinant of the matrix $\mathsf{M}$ is $[\nu_{2|1}^*-(2-\qq)\zeta^*][\nu_{2|1}^*-2(1-\qq)\zeta^*]$, so that the divergence of $b_3$--$b_8$ is again due that of $\kappa^*$ and $\mu^*$:
\beq
\alpha_\thr^{(b_3\text{--}b_8)}=\frac{4-d}{3d}.
\label{C5}
\eeq
The regions of divergence of the coefficients for $d=2$ and $d=3$ are depicted in Fig.\ \ref{fig3}.

\bibliographystyle{apsrev}

\bibliography{D:/Dropbox/Public/bib_files/Granular}

\end{document}